\documentclass[prb,aps,twocolumn,floatfix]{revtex4}

\usepackage{graphicx}
\usepackage{amsmath}
\usepackage{epstopdf}
\usepackage{appendix}
\usepackage{ulem}
\usepackage{color}
\usepackage{bm}

\newcommand{\beq}{\begin{equation}}
\newcommand{\eeq}{\end{equation}}
\newcommand{\bea}{\begin{eqnarray}}
\newcommand{\eea}{\end{eqnarray}}
\def\ang{\,\textrm{\AA}}

\newcommand{\avg}[1]{\left< #1 \right>} 

\def\eV{\,\textrm{eV}}
\def\T{\,\textrm{T}}
\def\kHz{\,\textrm{kHz}}
\def\MHz{\,\textrm{MHz}}
\def\ms{\,\textrm{ms}}
\def\mus{\,\mu\textrm{s}}

\def\Tr{\mathop{\textrm{Tr}}}
\def\peak{\textrm{peak}}

\begin{document}

\title{Spin decoherence in VOPc@graphene nanoribbon complexes}
\author{Xiao Chen$^{1,2}$, James N. Fry$^{1}$  and H. P. Cheng$^{1,2,3}$}
\affiliation{$^1$Department of Physics,  U. Florida, Gainesville FL 32611 USA}
\affiliation {$^2$Quantum Theory Project, University of Florida, Gainesville, FL 32611, USA}
\affiliation{$^3$Center for Molecular Magnetic Quantum Materials, University of Florida, Gainesville, FL 32611, USA}

\begin{abstract}

Carbon nanoribbon or nanographene qubit arrays can facilitate quantum-to-quantum transduction between light, charge, and spin, making them an excellent testbed for fundamental science in quantum coherent systems and for the construction of higher-level qubit circuits.  In this work, we study spin decoherence due to coupling with a surrounding nuclear spin bath of an electronic molecular spin of a vanadyl phthalocyanine (VOPc) molecule integrated on an armchair-edged graphene nanoribbon (GNR). Density functional theory (DFT) is used to obtain ground state atomic configurations. Decay of spin coherence in Hahn echo experiments is then simulated using the cluster correlation expansion method with a spin Hamiltonian involving hyperfine and electric field gradient tensors calculated from DFT. We find that the decoherence time $T_2$ is anisotropic with respect to magnetic field orientation and determined only by the hydrogen nuclear spins both on VOPc and GNR. Large electron spin echo envelope modulation (ESEEM) due to nitrogen and vanadium nuclear spins is present at specific field ranges and can be completely suppressed by tuning the magnetic field. The relation between these field ranges and the hyperfine interactions is analyzed. 
The effects of interactions with the nuclear quadrupole moments are also studied, validating the applicability and limitations of the spin Hamiltonian when they are disregarded.
\end{abstract}
\maketitle{}

\section{Introduction}

Synthesis of smooth-edge carbon nanoribbons (CNR) was first reported in 2008 \cite{WOS:1}. Soon after that, room temperature bottom-up fabrication techniques allowed ribbon growth with atomic precision \cite{WOS:2}. Over more than a decade, much effort was made to engineer electronic properties of CNR by modifying the edge states \cite{WOS:3, WOS:4, WOS:7, WOS:8}.
In 2016, Li \textit{et al.}\cite{WOS:5} improvised an efficient bottom-up procedure to synthesize armchair CNRs from molecular precursors via a polymerization method.
Later, Slota \textit{et al.}\cite{WOS:6} studied coherence control using graphene ribbons with magnetic edges realized by stable spin-bearing radical groups. 
In their system, long range
magnetic exchange coupling was observed, and spin coupling pathways were analysed from multi-frequency electron spin resonance. This work suggested that one might be able to attach magnetic molecules to nanoribbons and create a stable quasi-1D spin array. For qubit applications, one of the desirable features of a molecular qubit is to have spins localized in an individual molecule, from which well defined spin dimers (two qubits), trimers (three qubits), and chains of spins (qubit arrays) with coupling between the qubits can be constructed. 
Magnetic molecules such as vanadyl phthalocyanine (VOPc) are believed to be promising spin qubit candidates that can be used in quantum information sciences\cite{gaita2019molecular}. 
One of the competitive advantages of magnetic molecules compared to other solid-state spin qubits such as NV centers in diamond\cite{bar2013solid,schirhagl2014nitrogen} and phosphorus impurities in silicon\cite{steger2012quantum} is that the properties of molecular spins can be flexibly engineered in synthetic chemistry by choosing various metal centers and modifying peripheral ligands.
In addition, molecules are naturally monodisperse and molecular arrays allow qubits to follow a much more ordered lattice than defects in crystalline materials.

\begin{figure}[htp]
$\begin{array}{c}
\includegraphics[width=0.9\columnwidth]{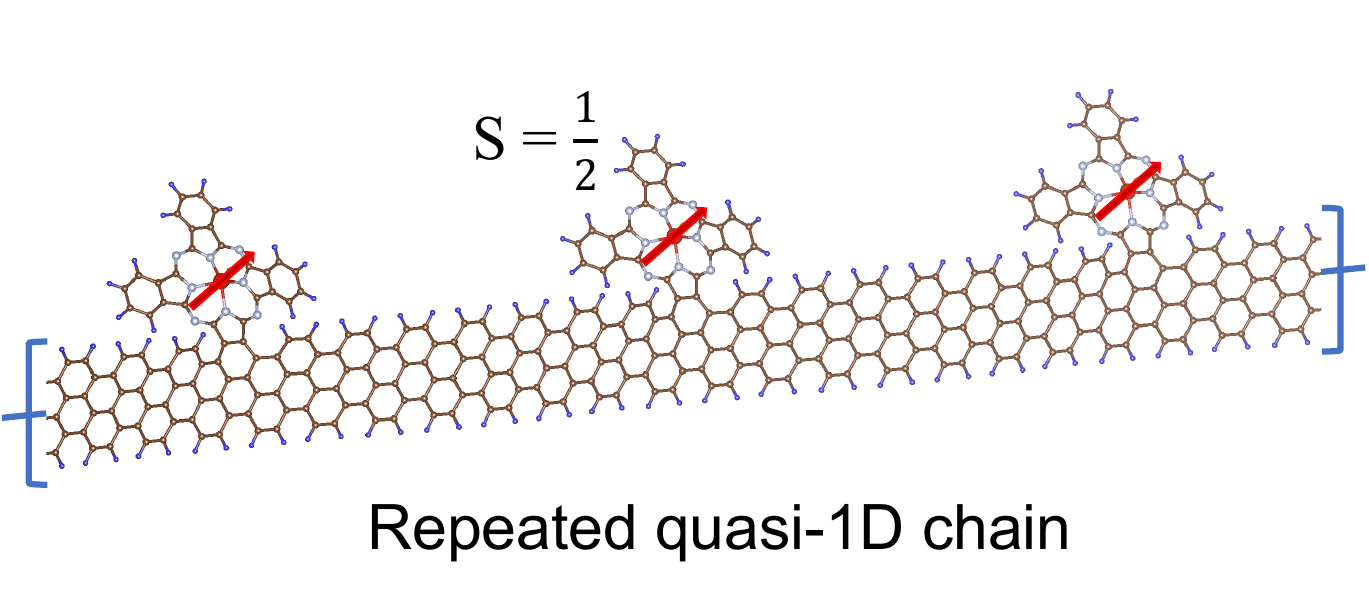}
\end{array}$\caption{VOPc integrated periodically onto graphene nanoribbons.
}\label{1d_array}
\end{figure}

To make use of magnetic molecular spins as qubits in quantum computing, one must find a way to create tunable and controllable couplings between different molecular spins\cite{gaita2019molecular,atzori2019second} in order to realize multi-qubit gates\cite{nielsen2002quantum,barenco1995elementary}. Otherwise, universal quantum gates\cite{nielsen2002quantum}, a set of quantum gates capable of creating entanglement between qubits that can serve as the building blocks from which any quantum gate can be constructed, are not possible. As part of the effort to fulfill this requirement, a first step is to find a microscopic structure such as the molecular spin chain above that can stably hold these magnetic molecules. Recently it was proposed that VOPc molecules can be integrated onto graphene nanoribbons with a structure as shown in Fig.~\ref{1d_array}. A recent development in synthetic chemistry has made the synthesis of this architecture possible with atomically precise control of the total length of the repeated one dimensional structure and the spacing between VOPcs\cite{li2018modular,yin2022programmable}. Qubit arrays anchored on (we denote as @) nanographene \cite{WOS:6,yoon2020liquid} facilitate quantum-to-quantum transduction between light, charge, and spin, making them an excellent testbed for fundamental science in quantum coherent systems and for the construction of higher-level qubit circuits. The coupling and
coherence between the molecular qubits can be controlled by varying the nanographene size, length, and edge
sites. Multiple-molecules@nanographene qubit arrays can then be integrated into scalable optical, microwave, and
electrical architectures to construct functional qubit circuits. These qubit@nanographene arrays can also be left
free in a solution or a gas, where they can be functionalized to bind to specific target sites for biological or
chemical sensing\cite{yu2021molecular}.

Qubit (or spin, for spin qubits) decoherence\cite{zurek2003decoherence}, the loss of relative phase between a superposition of qubit states that causes loss of quantum information, is one of the main constraining factors for the realization of quantum computing, or in general, quantum information sciences.
Therefore, properties of decoherence need to be studied when proposing any quantum computing architecture. For molecular spin qubits, the decoherence time $T_2$ contributed by dynamical quantum noises from the environment is conventionally measured in Hahn-echo experiments\cite{hahn1950spin,yang2016quantum,stoll2018epr}. Hahn echo is the simplest example of dynamical decoupling (DD). The key idea of the DD approach is to decouple a qubit from the source that causes decoherence by dynamically averaging out the noise from the environment via frequently flipping the spin\cite{yang2016quantum}. DD is a powerful method for suppressing the spin decoherence and has been widely proposed for quantum computing\cite{viola1998dynamical,ban1998photon,zanardi1999symmetrizing,viola1999dynamical}. In order to compare with experiments and considering the promising application of DD, the study of qubit decoherence usually includes a scenario with DD present.\cite{yang2016quantum,yang2008quantum,yang2009quantum,chen2020decoherence,kundu2023electron}

Inspired by promising advances in experiments and a lack of theoretical understanding for real-life material, as a first step, in this paper, we study qubit (spin) decoherence in the system of a single VOPc molecule integrated onto an infinitely long graphene nanoribbon (VOPc@GNR).
We focus on the pure dephasing regime where spin decoherence is due to the coupling of the molecular electron spin qubit to nuclear spins in the environment, which are the main decoherence source at low temperature where spin-phonon relaxation is suppressed. 
We use a combination of density functional theory (DFT) and the reduced density matrix (RDM) method to optimize atomic configurations, calculate electronic structure and parameters needed for spin Hamiltonians, and simulate the spin decoherence in Hahn-echo experiments. 
We report our results from simulations, compare them with a simple analytical product rule model, and discuss underlying physics. The paper is organized as follows: 
in Sect. II, we describe the first-principles methods we use to obtain interactions for the spin Hamiltonian and the cluster correlation expansion method used to calculate the coherence function.
In Sect. III, we report our results for the Hahn-echo coherence functions and discuss the implications of and reasons for certain physical behaviors. This section includes atomic and electronic structures, spin decoherence, two-spin model studies, and effects of nuclear quadrupole moments. 
Final conclusions are in section IV.

\section{Method}\label{sec:GL}

We use the Vienna Ab initio Simulation Package (VASP)\cite{kresse1993ab,kresse1996efficiency,kresse1996efficient} to perform Density Functional Theory (DFT) calculations of the atomic configuration and electronic structure of the VOPc@GNR system. The infinitely long graphene nanoribbon (GNR) we consider is the 6-AGNR as in Fig.~\ref{1d_array}, which has six rows of carbon atoms with armchair edges, a GNR recently synthesized with atomic precision\cite{li2018modular,yoon2020liquid}.
The rectangular simulation cell is chosen to have dimensions of 
$51.66 \times 35 \times 20 \ang$ along the $x$-, $y$- and $z$-directions, where the quasi-one-dimensional GNR is along $x$. These dimensions ensure that the part of the GNR in the cell far from the attached VOPc molecule approaches the structure of free GNR and that vacuum in $y$- and $z$-directions between the VOPc@GNR system and its periodic images is adequately thick. Given the large cell size, first Brillouin zone 
integrations are done using a $\Gamma$ point $k$-point grid. 
The energy cutoff of the plane-wave basis was set to $600 \eV$ throughout all DFT calculations. The total energy tolerance for electronic self-consistency and the force tolerance for ionic relaxation are set to $1 \times 10^{-8} \eV$ and $0.01\eV/\textrm{\AA}$, respectively. During ionic relaxation, we adopt the Perdew-Burke-Ernzerhof (PBE) 
exchange correlation energy functional\cite{perdew1996generalized}.
After ionic relaxation is complete we use the PBE+$U$ method for electronic structure calculations to account for localization of electrons on the vanadium atom, with Hubbard $U$ set to $4.3\eV$, according to previous linear response calculations in the literature\cite{Hubbard_U}.
In the simulation of spin decoherence, the GNR in the relaxed atomic configuration in the DFT simulation cell is extended on both sides to infinity in the $x$-direction with the free structure of the GNR.
The hyperfine interaction and electric field gradient (EFG), which are useful in constructing spin Hamiltonians, are calculated using built-in routines of VASP, where the hyperfine interaction tensor for a nuclear spin $\bm{I}$ at position $\bm{R}_{I}$ is the sum of Fermi contact and dipolar terms,
which in Cartesian components are given by  
\beq
(A_{FC}^{I})_{ij}=\frac{2}{3}\frac{\mu_{0}\gamma_{e}\gamma_{I}}{\avg{Sz}}\delta_{ij}\int\delta_{T}(\bm{r})\rho_{s}(\bm{r}+\bm{R}_{I}) \, d\bm{r} \label{Fermi_contact_hyperfine}
\eeq
and
\beq
(A_{D}^{I})_{ij}=\frac{\mu_{0}}{4\pi}\frac{\gamma_{e}\gamma_{I}}{\avg{S_{z}}} \int\frac{\rho_{s}(\bm{r}+\bm{R}_{I})}{r^{3}} \, \frac{3r_{i}r_{j}-\delta_{ij}r^{2}}{r^{2}}\, d\bm{r} . \label{dipolar_hyperfine}
\eeq
In these expressions, $r_i$ are the components of the position vector $\bm{r}$, with $r=|\bm{r}|$, $\rho_s$ is the spin density, $\mu_0$ the magnetic susceptibility of free space, $\gamma_e$ the electron gyromagnetic ratio, $\gamma_I$ the nuclear gyromagnetic ratio of the nucleus, and $\avg{S_{z}}$  the expectation value of the $z$-component of the electronic spin. $\delta_T(\bm{r})$ is a smeared $\delta$ function, as described in the Appendix of Ref.~[\onlinecite{blochl2000first}].
The core contribution\cite{yazyev2005core} to the Fermi contact part of the hyperfine interaction has been included in computing $A_{FC}$ in eq.~(\ref{Fermi_contact_hyperfine}).
We use the hyperfine interaction tensor calculated from DFT only for the nuclear spins on the VOPc molecule, since the electron spin density is highly localized around the vanadium center on the VOPc molecule. The nuclear spins on the GNR are from hydrogen nuclei (C nuclei are purely spinless $^{12}\textrm{C}$ due to the $1.1\%$ low natural abundance of spin-$1/2\,$ $^{13}\textrm{C}$), at a greater distance from the molecular electronic spin than those in VOPc.
For the hyperfine interaction of these hydrogen nuclei, the magnetic point dipole-dipole interaction was adopted.
We note that spin-orbit coupling is not included in our DFT calculations unless explicitly mentioned.

Calculations of coherence functions of the VOPc molecular spin in VOPc@GNR are conducted using the cluster correlation expansion (CCE) method\cite{yang2008quantum,yang2009quantum}, as implemented in the PyCCE code\cite{onizhuk2021pycce}. 
The spin Hamiltonian describing the dynamics of the electronic spin-${1}/{2}$ center, which we will call the central spin, interacting with a bath of nuclear spins in the presence of an external magnetic field takes the form
\beq
H=H_{S}+H_{B}+H_{SB} , \label{spin_Hamiltonian}
\eeq
where the terms describing the central spin and its interactions with the surrounding bath nuclear spins are 
\bea
H_{S}&=&-\gamma_{e}\bm{B}\cdot\hat{\bm{S}},\nonumber\\
H_{SB}&=&\sum_{i}\hat{\bm{S}}\cdot\mathbf{A}_{i}\cdot\hat{\bm{I}}_{i} .
\eea
The gyromagnetic ratio of the central spin $\gamma_e$ is assumed to take the value for a free electron. The hyperfine interaction tensors $\mathbf{A}_{i}$ are calculated as described
above. The VOPc@GNR system we study includes nuclear spin operators ${\bm{I}}_{i}$  for one spin-${{7}/{2}}$ vanadium (V) nucleus and eight spin-${1}$ nitrogen (N) nuclei as well as all nuclear spin-${{1}/{2}}$ hydrogen (H).
The Hamiltonian for the bath is given by
\beq
H_{B}=-\sum_{i}\gamma_{i}\bm{B}\cdot\hat{\bm{I}}_{i}+\sum_{i}\hat{\bm{I}}_{i}\cdot\mathbf{P}_{i}\cdot\hat{\bm{I}}_{i}+\sum_{i<j}\hat{\bm{I}}_{i}\cdot\mathbf{J}_{ij}\cdot\hat{\bm{I}}_{j} . \label{bath}
\eeq
The first term is the Zeeman energy with $\gamma_i$ the gyromagnetic ratio of nuclear spin $i$. The second term is the nuclear quadrupole interaction (NQI), which is present only for nuclear spins with spin quantum number larger than one half, here the V and N nuclear spins in VOPc@GNR. The quadrupole interaction tensor $\mathbf{P}_{i}$ is computed from the EFG tensor obtained from DFT. The third term is the magnetic point dipolar interaction between nuclear bath spins.


The combined
central spin and bath system is initially prepared in a product state of
the form
\beq
\hat{\rho}(0)=\hat{\rho}_{S}(0)\otimes\hat{\rho}_{B}(0),\label{eq:initial_total_dm}
\eeq
where $\hat{\rho}_{S}(0)$ and $\hat{\rho}_{B}(0)$ are reduced density operators at $t=0$ of the central spin and the bath, respectively. $\hat{\rho}_{S}$ represents a pure state of an equal superposition of qubit states chosen as the two central spin eigenstates of $H_{S}$, 
\beq
\hat{\rho}_{S}(0) = |\psi\rangle\langle\psi|, \qquad
|\psi\rangle = \frac{1}{\sqrt{2}}(|0\rangle+|1\rangle), \label{initial_S_state}
\eeq
where $|0\rangle$ and $|1\rangle$ are $m_{s}=+\frac{1}{2}$ and $m_{s}=-\frac{1}{2}$ eigenstates of the central spin, and $\rho_{B}(0)$ is a product state of reduced density operators of individual nuclear spins, 
\beq
\hat{\rho}_{B}(0)=\otimes_{i}\hat{\rho}_{i}, \label{initial_B_state}
\eeq
with each nuclear spin assumed to be purely random:  
$\hat{\rho}_{i}=\hat{I_0}/(2I+1)$, where $I$ is its spin quantum number and $I_0$ the identity operator. This assumption is justified, as we will consider temperatures much larger than the nuclear spin Zeeman energies, which are on the order of $10^{-5}$--$10^{-2} \, \textrm{K}$.

Spin decoherence is studied by simulating the time dependence of the normalized coherence function of the central spin, defined by the off-diagonal elements of its reduced density matrix, as in a Hahn-echo experiment, 
\beq
L(t=2\tau)=\frac{\langle1|\hat{\rho}_{S}(t)|0\rangle}{\langle1|\hat{\rho}_{S}(0)|0\rangle},\label{coherence_function}
\eeq
where $\tau$ is the pulse delay time. The decay of the coherence function, or decoherence, is a loss of information on the relative phase between the two qubit states in superposition and a breakdown of the superposition itself. The lifetime of this decay in Hahn-echo experiments is commonly called the Hahn-echo $T_2$ or simply $T_2$ in the literature.


For a large bath of a few hundred spins or more, the coherence function
can be efficiently calculated with the CCE method.
The key idea of the CCE is that the decoherence of a central electron spin due to interaction with a nuclear spin bath can be exactly expanded as a product
of contributions from irreducible correlations of bath-spin clusters\cite{yang2008quantum,yang2009quantum,onizhuk2021pycce},
\beq
L(t)=\tilde{L}_{\left\{ \emptyset\right\}}(t)\prod_{\left\{ i\right\}}\tilde{L}_{\left\{ i\right\}}(t)\prod_{\left\{ ij\right\}}\tilde{L}_{\left\{ ij\right\}}(t)\prod_{\left\{ ijk\right\}}\tilde{L}_{\left\{ ijk\right\}}(t) \dots , \label{CCE_expansion}
\eeq
where $\tilde{L}_{\left\{ \emptyset\right\}}(t)$ is the phase factor of the free evolution of the central spin, $\tilde{L}_{\left\{ i\right\}}(t)$ is the contribution from single bath spin $i$, $\tilde{L}_{\left\{ ij\right\}}(t)$ is the contribution from unordered spin pairs $\left\{ij\right\}$, and $\tilde{L}_{\left\{ ijk\right\}}(t)$ from a cluster of three different spins, \textit{etc.} The irreducible correlation of a cluster is defined iteratively as\cite{yang2008quantum,yang2009quantum,onizhuk2021pycce}
\beq
\tilde{L}_{C}=\frac{L_{C}}
{\displaystyle\prod_{C'\subset C}\tilde{L}_{C'}} , \label{def_of_LC}
\eeq
where ${L}_{C}$ is the coherence function of the central spin if only the terms in the spin Hamiltonian (\ref{spin_Hamiltonian}) containing the central spin $\hat{\bm{S}}$ and bath spins $\hat{\bm{I}}_{i}$ in cluster $C$, but no other bath spins, are present. We label the sum of these terms as $H_{C+S}$. 
To simulate a Hahn-echo experiment,
\beq
L_{C}(t=2\tau)=\bigl\langle 0 \big| \, {\Tr}_{C} 
[\hat{U}_{C+S}(t) \, \hat{\rho}_{C+S} \, \hat{U}_{C+S}^{\dagger}(t)] \,  \big|1\bigr\rangle, \label{LC_Hahn_echo}
\eeq
\beq
\hat{U}_{C+S}(t)=e^{-i\hat{H}_{C+S}\tau}e^{-i\pi\hat{S}_{x}}e^{-i\hat{H}_{C+S}\tau}, 
\eeq
where $\hat{\rho}_{C+S}$ is the initial density matrix as a product state as in eqs.~(\ref{initial_S_state}) and (\ref{initial_B_state}) for the subsystem of the central spin and the bath-spin cluster $C$, $\Tr_{C}$ is the partial trace over the state space of $C$, and an ideal $\pi$ pulse flips the central spin at the pulse delay time $\tau$.
$\hbar$ has been set to $1$.
In the calculation of $L_{C}(t=2\tau)$, the conventional scheme\cite{onizhuk2021pycce},  which assumes no central spin flipping is adopted. This is valid in the problem of electron spin decoherence in a nuclear spin bath, since the electron Zeeman energy is three to four orders of magnitude larger than that of the nuclear spins under the same field.
Since contributions from subcluster correlations are divided from ${L}_{C}$, $\tilde{L}_{C}$ represents the irreducible correlation between all spins in $C$.

If the expansion $(\ref{CCE_expansion})$ converges rapidly, it is valid to truncate it in practice. The maximum number of spins in the clusters included in the expansion determines the order of the CCE approximation. For example, the CCE order 2 (CCE-2) expansion includes only and all irreducible cluster correlation contributions up to two-spin clusters.
The CCE expansion provides the essentially exact result if the expansion is truncated at an order where the coherence function is already convergent with respect to CCE order.
In the present work, we limit ourselves to CCE order 4 (CCE-4), as our convergence studies show there is essentially no change in $L(t=2\tau)$ when going from CCE-4 to CCE-5. Convergence test results are presented in Appendix \ref{ConvergenceTest}.

\section{Results and discussion}\label{sec:model}

In this section, we report our results for VOPc@GNR. The atomic configuration and electronic structure of the system are presented first, followed by the results for spin decoherence and related analyses. Lastly, we investigate the effect of NQI.

\subsection{Atomic configuration and electronic structure of VOPc@GNR}

\begin{figure}
\centering
$\begin{array}{c}
\includegraphics[width=0.38\textwidth]{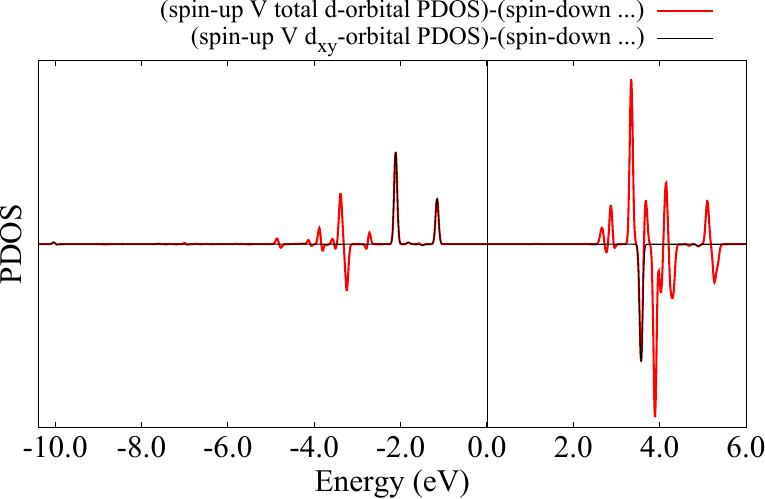}\\
\includegraphics[width=0.3\textwidth]{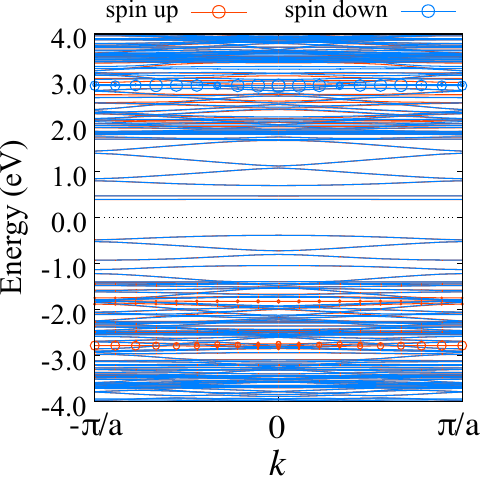}
\end{array}$

\caption{(Top) The difference between the spin-up and -down density of states of a single VOPc projected onto all V $d$-orbitals (red curve) and onto the V $d_{xy}$ orbital (black curve). (Bottom) Fat band analysis of the VOPc@GNR complex where the radius of any circle in the bands is proportional to the value of the $k$-resolved PDOS for the V $d_{xy}$ orbital at the energy $E$ and momentum $k$ of the position of the circle. The Fermi level is set to zero in both plots.}\label{DFT_results}

\end{figure}

We first report the results of DFT calculations atomic configuration and electronic structure of a single VOPc molecule and VOPc@GNR. Figure~\ref{DFT_results}~(Top) shows the difference between the spin-up and -down projected density of states (PDOS) summed over all V $d$ orbitals (red curve) and for the V $d_{xy}$ orbital (black curve) of an isolated single VOPc molecule. As anticipated, the spin density of a single VOPc molecule is contributed by an unpaired $d_{xy}$ electron. Figure~\ref{DFT_results}~(Bottom) shows the 
fat band analysis of the VOPc@GNR system, which we will return to later. For an isolated VOPc molecule, we also calculate energy as a function of spin orientation with inclusion of spin-orbit coupling. The preferred spin direction is along the vanadium-oxygen (V-O) bond and the energy of this spin direction is $ 41 \,\textrm{$\mu$eV} $ lower than the situation in which the spin is perpendicular to the V-O bond or in the plane of the molecule. 

\begin{figure*}[htp]
$\begin{array}{ccc}
\includegraphics[width=0.32\linewidth]{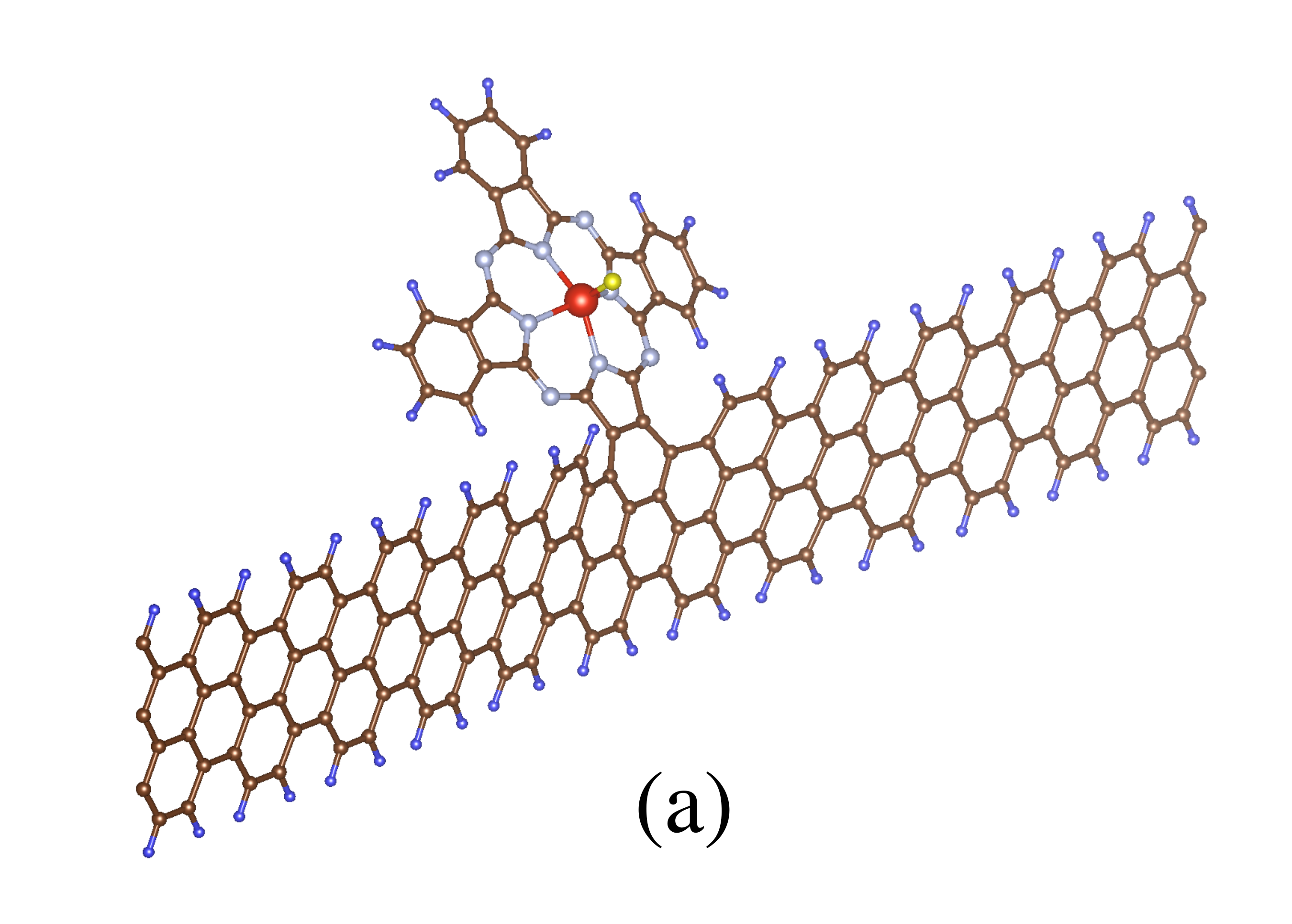}&
\includegraphics[width=0.32\linewidth]{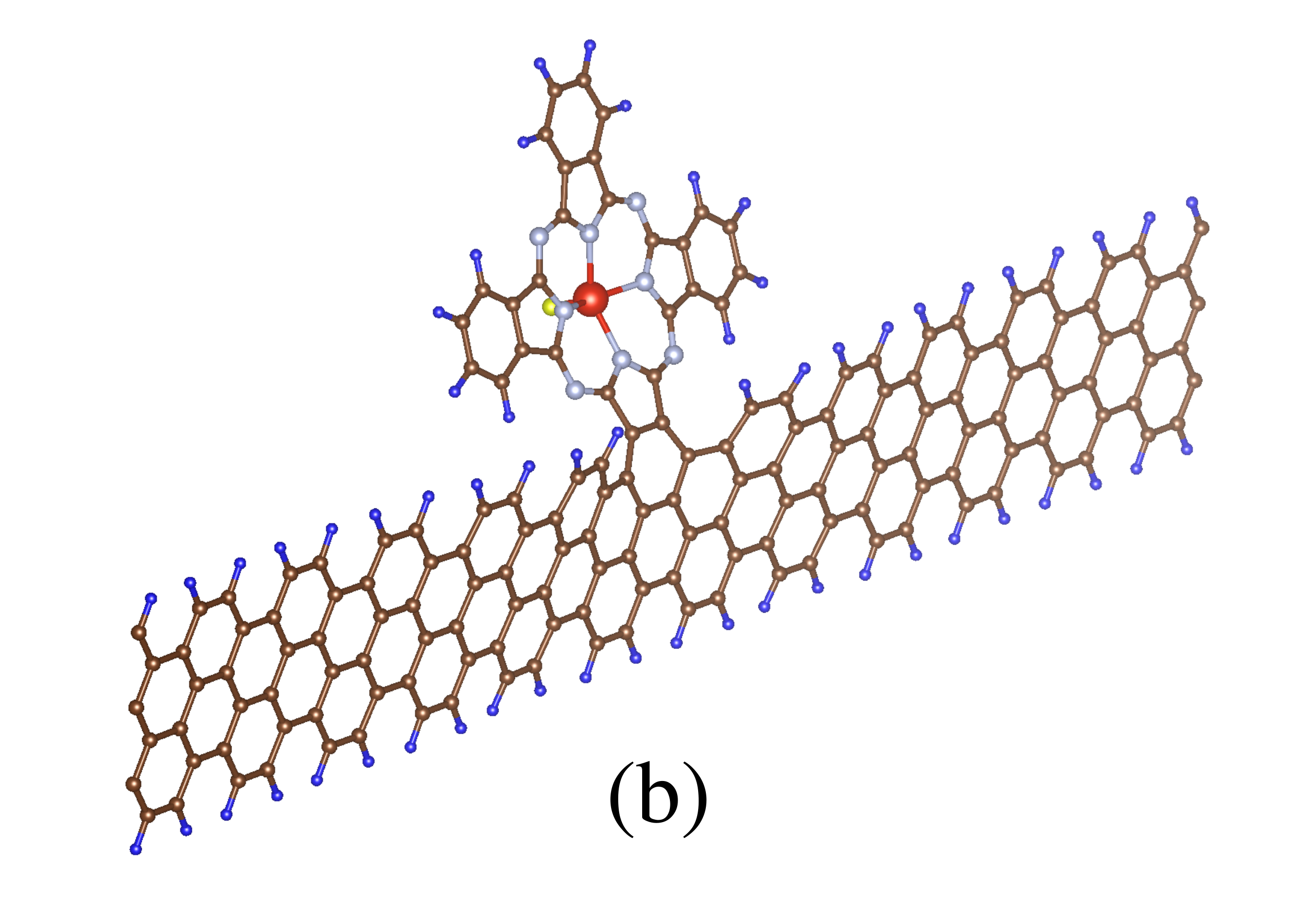}&
\includegraphics[width=0.32\linewidth]{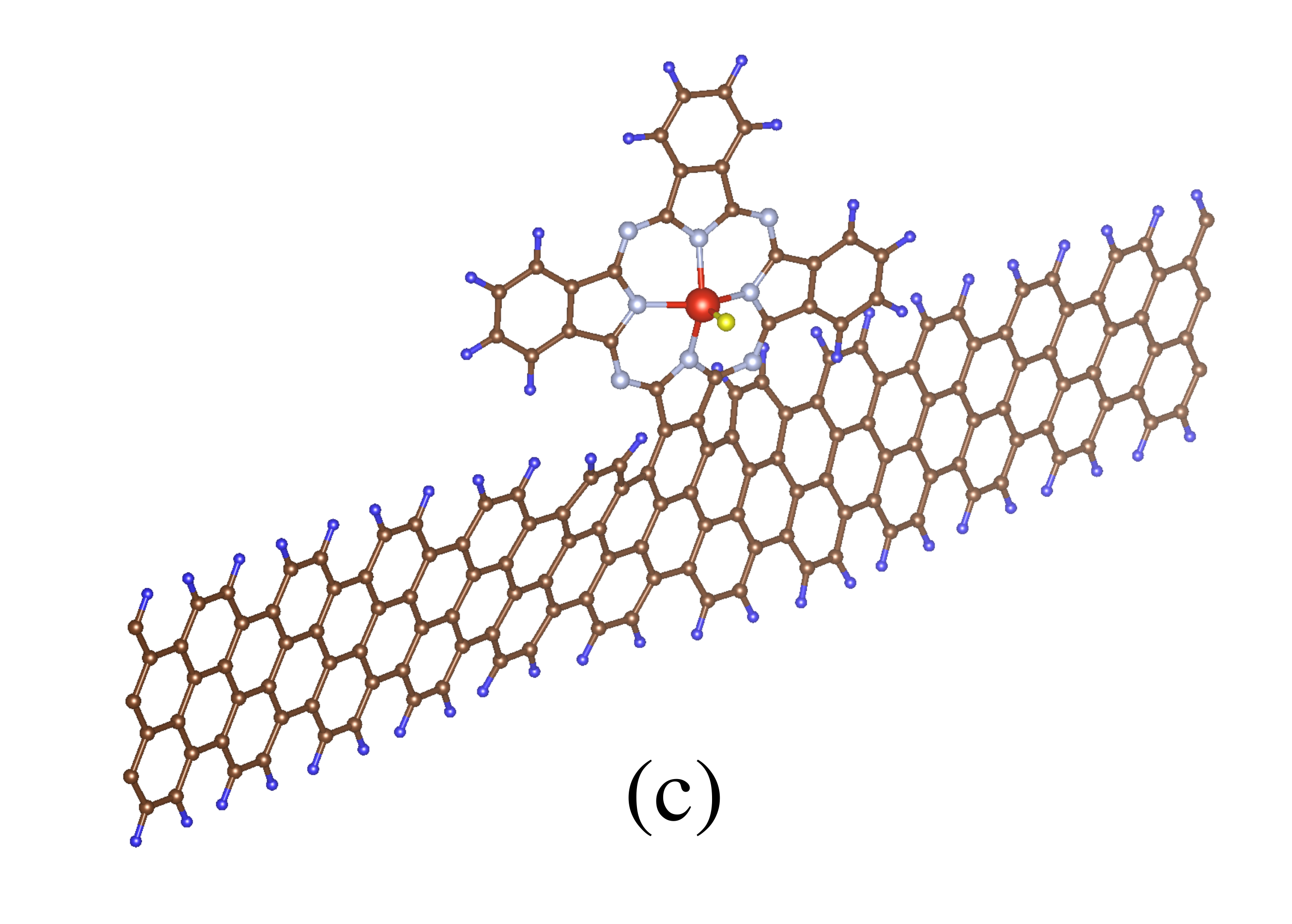}
\end{array}$
\caption{Isomeric structures of the VOPc@GNR system: {(a)} Oxygen near GNR plane; {(b)} Oxygen off GNR plane; {(c)} twisted axis configuration. Color code: red for vanadium, yellow for oxygen, white for nitrogen, brown for carbon, and blue for hydrogen.}\label{fig:isomeric_structures}
\end{figure*}

When a VOPc molecule is integrated onto an armchair-edged GNR with a width of two honeycomb units, three isomeric structures are possible, as shown in Fig.~\ref{fig:isomeric_structures}. The energies of these three configurations in Fig.~\ref{fig:isomeric_structures}(a), \ref{fig:isomeric_structures}(b) and \ref{fig:isomeric_structures}(c) are 0.0, 
$ 106, \textrm{meV} $, and $ 232 \, \textrm{meV} $, respectively, with the ground state energy being set to zero. In all cases, the plane of the phthalocyanine (Pc) ligand deviates significantly from that of the GNR due to a strong repulsion between hydrogen atoms on the ligand near the GNR and those on the GNR near the ligand. The result shows that if the repulsive force pushes both horizontal isoindole units to one side of the GNR, structures in \ref{fig:isomeric_structures}(a) and \ref{fig:isomeric_structures}(b) are realized with the oxygen atom nearer to the plane of GNR in \ref{fig:isomeric_structures}(a). If the repulsion pushes two horizontal isoindole units to different sides of the GNR then we obtain the structure in \ref{fig:isomeric_structures}(c), where the VOPc molecule is twisted with respect to the GNR. 
In the rest of the paper, we focus on the ground state of the three configurations, as shown in \ref{fig:isomeric_structures}(a). 

We calculate the DFT band structure of VOPc@GNR and the $k$-resolved PDOS, \textit{i.e.} we perform the fat band analysis [see Fig.~\ref{DFT_results}~(Bottom) for band structure and $k$-resolved density of states projected onto the V $d_{xy}$ orbital]. 
It is found that in this VOPc@GNR complex a localized molecular spin-${1}/{2}$ is still contributed by the unpaired $d_{xy}$ electron on the V atom.

\begin{figure}[htp]
$\begin{array}{c}
\includegraphics[width=0.7\columnwidth]{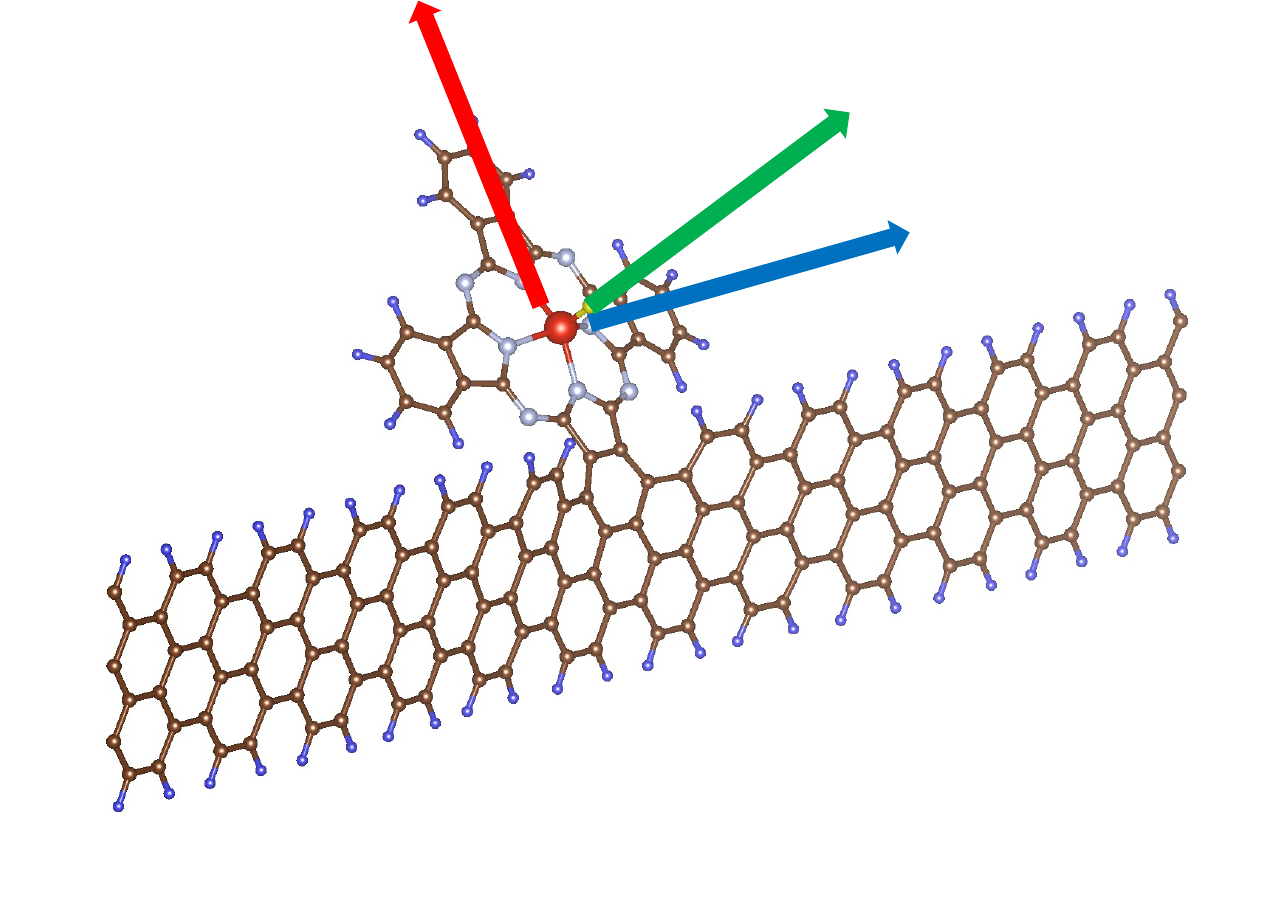}
\end{array}$\caption{The three mutually perpendicular directions of magnetic field considered are indicated by the arrows. (i) Blue arrow: the GNR direction ($x$-direction). (ii) Red arrow: the direction perpendicular to both the GNR direction and the V-O bond. (iii) Green arrow: the direction of the V-O bond.
}\label{fig:field_directions}
\end{figure}

\begin{figure*}[htp]
$\begin{array}{ccc}
\includegraphics[width=0.31\linewidth]{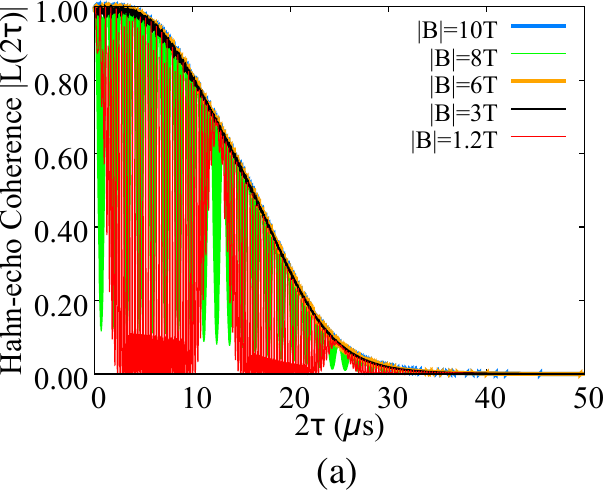}&
\includegraphics[width=0.31\linewidth]{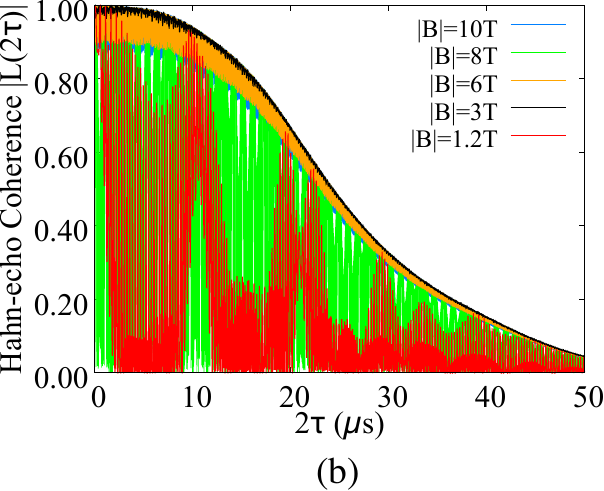}&
\includegraphics[width=0.31\linewidth]{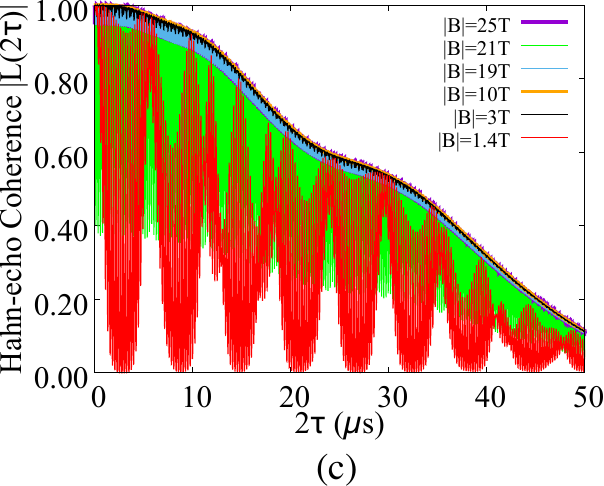}
\end{array}$
\caption{Dependence of the Hahn echo coherence function on the magnetic field direction and strength: {(a)} magnetic field is in the direction along the GNR; {(b)} magnetic field is in the direction (ii) in Fig.~\ref{fig:field_directions} and as described in the main text; {(c)} magnetic field is in the direction parallel to the V-O bond. }\label{fig:field_direction_and_strength_dependence}
\end{figure*}

\subsection{Spin decoherence}


Spin decoherence in the VOPc@GNR system is contributed by the nuclear spins within a radius of roughly $26\ang$ from the central spin. This is discovered by increasing the radius of a sphere centered at the position of the central spin at the V atom, and for each radius, including only the spins of the nuclei inside the sphere as the bath spins in the spin Hamiltonian. Decay of the Hahn-echo coherence function stops exhibiting any further change when this bath radius is increased to $26\ang$ [See Appendix \ref{ConvergenceTest} and Fig.~\ref{fig:convergence}(a)].
This test of tuning the bath radius also shows why the decoherence rate is much faster in VOPc@GNR [black curve in Fig.~\ref{fig:convergence}(b), for which $T_2=17.4\mus$] compared to that in a single VOPc molecule, which basically corresponds to setting the bath radius to only $8\ang$ in VOPc@GNR [green curve in Fig.~\ref{fig:convergence}(b), which does not show any sign of decay at least up to $50\mus$], indicating the important role played by the H nuclear spins on the GNR on central spin decoherence.

After convergence tests on the coherence function are completed  (Appendix \ref{ConvergenceTest}), we start with investigations of the dependence of spin decoherence in the VOPc@GNR system on magnetic field direction and strength. Three mutually perpendicular directions were considered, as shown in Fig.~\ref{fig:field_directions}. 
These are (i) parallel to the GNR direction ($x$-direction, blue arrow), (ii) in a direction perpendicular to both the GNR direction and the V-O bond (red arrow, approximately parallel to the isoindole unit oriented perpendicularly to the $x$-direction), and (iii) in the direction of the V-O bond (green arrow).
Without losing generality, we first show results excluding the NQI, \textit{i.e.} the second term in Eq.~\ref{bath}, which is dropped for now; we will discuss the effect of including NQI in the Hamiltonian in Sect.~\ref{SubSect:NQI_effect}. The Hahn-echo coherence functions for field directions (i),(ii) and (iii) are shown in Fig.~\ref{fig:field_direction_and_strength_dependence}(a), (b) and (c), respectively, as function of pulse delay time $2\tau$ as computed according to 
Eqs.~(\ref{coherence_function})--%
(\ref{LC_Hahn_echo}).
Here we note that all Hahn-echo coherence functions $L$ calculated for VOPc@GNR have negligble imaginary part, and we will show the magnitude of $L$ as the correct measure.
We see that the spin decoherence depends on the relative orientation of the single VOPc molecule and the direction of the magnetic field, due to anisotropic hyperfine and inter-nuclear-spin interactions. One also observes that 1) the envelope of the Hahn-echo coherence function decay depends on the direction of the field but not its strength and 2) The oscillation under the envelope, know as electron spin echo envelope modulation (ESEEM), does depend on field strength.
This envelope has the longest $T_2$ of $34.2 \mus$ when the field is along the V-O bond, direction (iii), followed by a field in direction (ii) with a $T_2$ of $28.3\mus$, and then by field along the GNR with a $T_2$ of $17.4\mus$.
The value of $T_2$ is obtained by fitting the coherence function to a stretched exponential $\textrm{exp}[-(t/T_2)^n]$.  In our analysis later in the paper we show that the envelope is determined only by the H nuclear spins in the system. 
Interestingly, for both fields along the GNR and along direction (ii), at relatively small field strengths of around 0--$2\T$ we observe a large ESEEM effect as represented by the red curves in Fig.~\ref{fig:field_direction_and_strength_dependence}(a) and (b). When we increase the field strength to 3--$6\T$, the ESEEM is suppressed, while further increasing the field to around 7--$9\T$ a large ESEEM appears again, as shown by the green curves. In the third field direction, which is along the V-O bond, similar ESEEM effects at relatively small fields of around 0--$2\T$ are observed, while those at large fields are only observed instead at around $20-24\T$ (Fig.~\ref{fig:field_direction_and_strength_dependence}(c)). 
In Hahn-echo experiments done on the VOPc molecules using pulsed electron paramagnetic resonance spectroscopy, large ESEEM is indeed found at small fields\cite{atzori2016room,follmer2020understanding}.

The large ESEEM rapidly and significantly modifies the coherence function. Although it still represents a theoretically tractable coherent process between few nuclear spins and the central spin, as we will see soon below, it is desirable to avoid them in potential realizations of spin qubits. A realization of a spin qubit working with dynamical decoupling pulses such as the Hahn-echo pulse sequence while the ESEEM effect is present must incorporate accurately the frequencies of the ESEEM so that one can predict when coherence function becomes close to one and the superposition state of the qubit is recovered, but these frequencies depend on the hyperfine interactions and nuclear Larmor frequencies, which are sensitive to perturbations in the local environment. 
One can tune the magnetic field strength to suppress ESEEM as observed above, therefore it is helpful to understand at what field strength large ESEEM can occur.
In the following analysis we try to understand what nuclear species give rise to the ESEEM observations and why only in the specific ranges of magnetic field strength found above. We take the field direction along GNR as an example but note that the qualitative results also hold for the other two directions. 



\begin{figure*}[htp]
$\begin{array}{cc}
\includegraphics[width=0.4\linewidth]{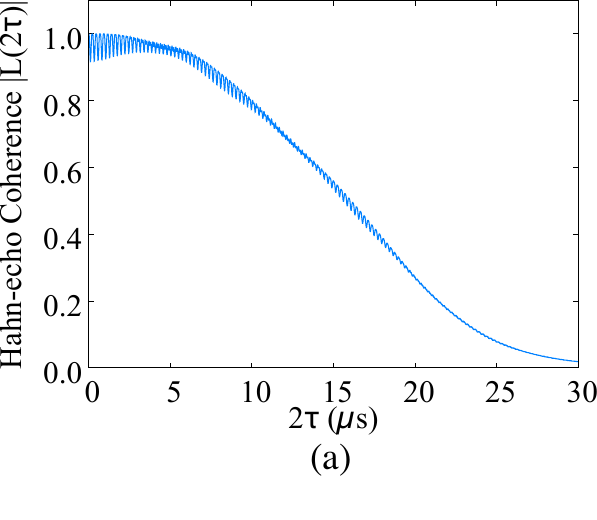}&
\includegraphics[width=0.4\linewidth]{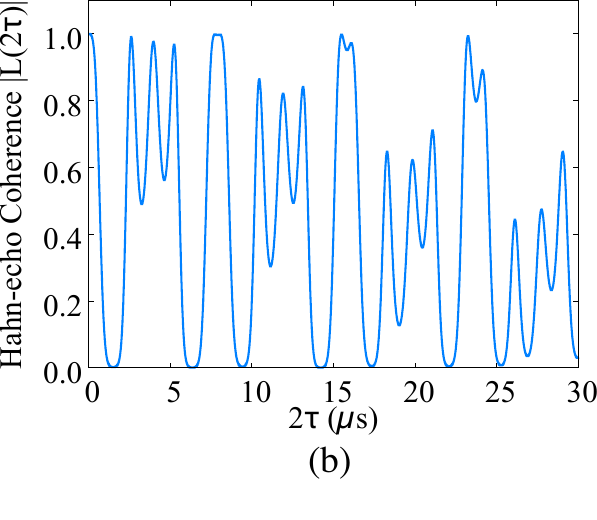}\\
\includegraphics[width=0.4\linewidth]{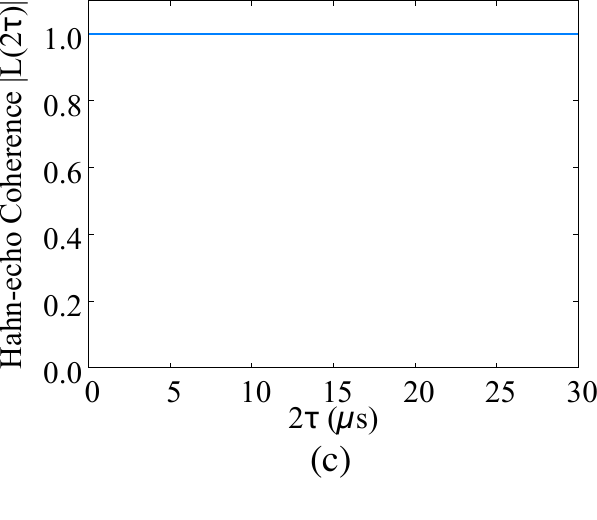}&
\includegraphics[width=0.4\linewidth]{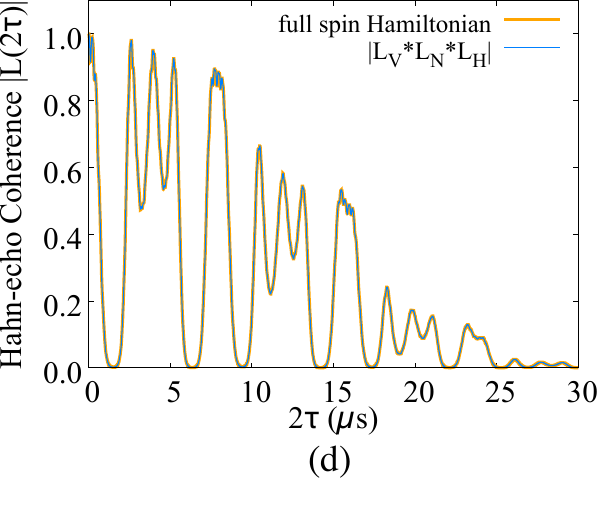}
\end{array}$
\caption{Product rule for the Hahn-echo coherence function among different elements in the VOPc@GNR system. A magnetic field of strength $0.2\T$ is applied in the direction along the GNR. The coherence function is contributed by the central spin coupling to {(a)} H nuclei, {(b)} N nuclei, and {(c)} the V nucleus. In {(d)}, the coherence function due to all elements present in the spin Hamiltonian is seen to closely follow the product of (a),(b) and (c). }\label{fig:product_elements_1a}
\end{figure*}

\begin{figure*}[htp]
$\begin{array}{cc}
\includegraphics[width=0.4\linewidth]{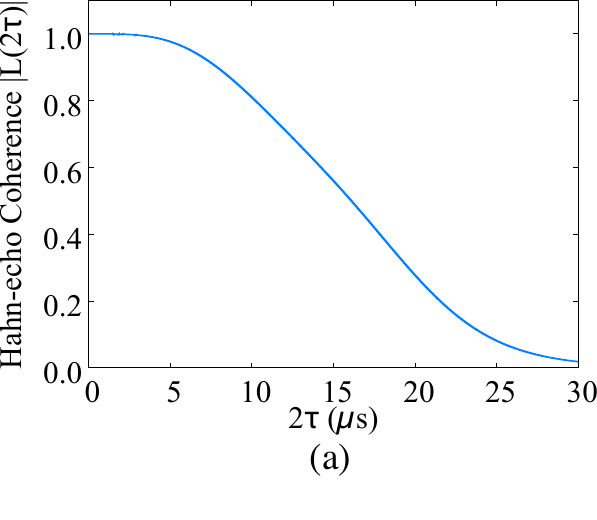}&
\includegraphics[width=0.4\linewidth]{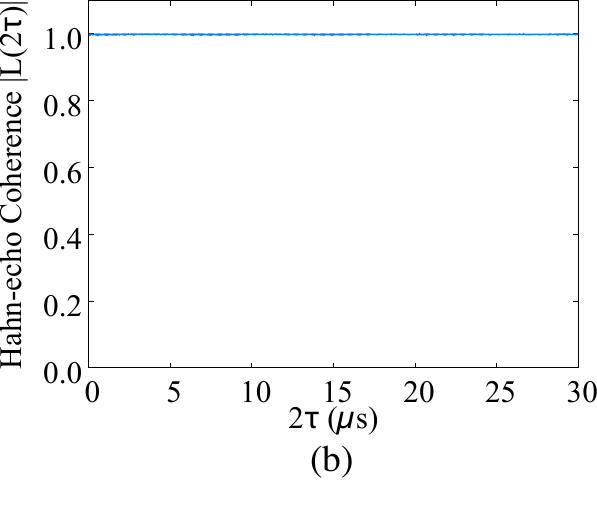}\\
\includegraphics[width=0.4\linewidth]{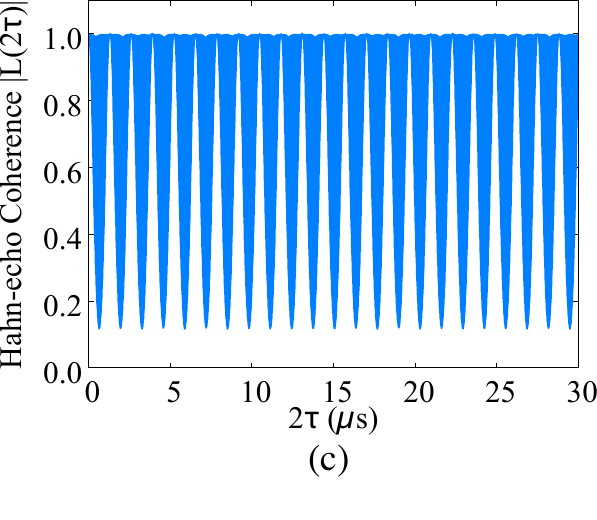}&
\includegraphics[width=0.4\linewidth]{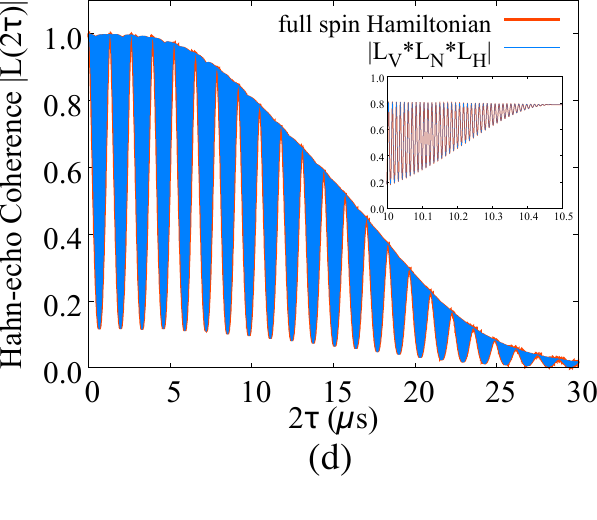}
\end{array}$
\caption{Product rule for the Hahn-echo coherence function among different elements in the VOPc@GNR system. A magnetic field of strength $8.0\T$ is applied in the direction along the GNR. The coherence function is contributed by the central spin coupling to {(a)} H nuclei, {(b)} N nuclei and {(c)} the V nucleus. In {(d)}, the coherence function due to all elements present in the spin Hamiltonian is seen to follow the product of (a),(b) and (c).
Inset is a zoom-in at short time scale showing the details of the oscillatory behavior and the close agreement between the two curves. }\label{fig:product_elements_1b}
\end{figure*}

\begin{figure}[htp]
$\begin{array}{c}
\includegraphics[width=0.95\columnwidth]{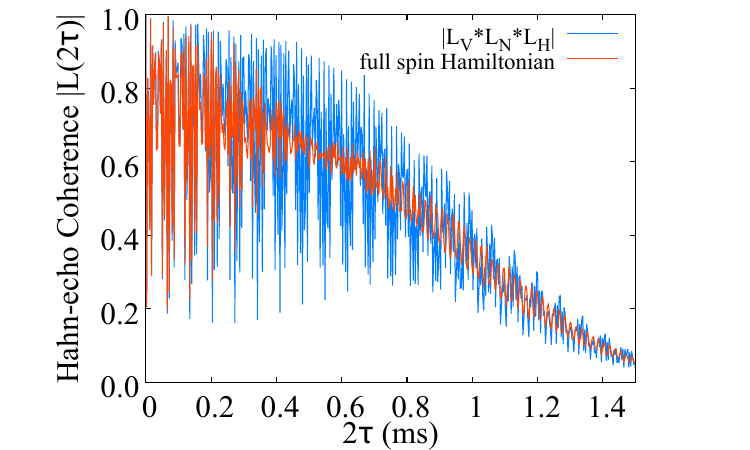}
\end{array}$\caption{Hahn-echo coherence function simulated with the same central spin, V and eight N nuclear spins as in VOPc@GNR immersed in a random sparse H nuclear bath.
\label{fig:comparison} }
\end{figure}

In order to determine what nuclear spin species give rise to the observed ESEEM effects, we compute the decoherence caused by each individual nuclear spin species separately, which is achieved by allowing the central spin coherence function to evolve with a spin Hamiltonian reduced from the full spin Hamiltonian by keeping only one species of nuclear spin at a time while dropping all the terms in Eq.~(\ref{spin_Hamiltonian}) containing nuclear spins $\hat{\bm{I}}_{i}$ of other species. From the results as represented by the data shown in Fig.~\ref{fig:product_elements_1a} and in Fig.~\ref{fig:product_elements_1b}, we learn that the ESEEM at a relatively small field strength of 0--$2\T$ is contributed by the N nuclear spins and in the range of 7--$9\T$ by the V nuclear spin. 
In each of Fig.~\ref{fig:product_elements_1a} and \ref{fig:product_elements_1b}, the coherence function when only H, N or V nuclear spins are present are in subplots (a), (b) and (c), respectively.
More interestingly, when we compute the product of the coherence functions due to all three nuclear spin species and compare the result with the coherence function as a result of the full spin Hamiltonian, we find a very good agreement with the product of the three, as can be seen from Figs.~\ref{fig:product_elements_1a}(d) and \ref{fig:product_elements_1b}(d) [Examples of this product rule for directions (ii) and (iii) are shown in Appendix~\ref{product_rules_directions_ii_and_iii}.]. This product rule is exact in the limit of strong field and when the interactions between different groups of nuclear bath spins, here different species, are set to zero.\cite{mims1972envelope,stoll2009general,schweiger2001principles} 
This has been observed previously in other systems and is due to suppression of spin flip-flop processes between different nuclear spin species because of a large discrepancy in nuclear Larmor frequencies.\cite{ye2019spin,seo2016quantum} 
With this product rule, it is obvious that the coherence function due to H nuclei provides the envelope of the ESEEM from the full Hamiltonian and so determines $T_2$.
Note that for a system where the decoherence time is large enough such that a correlation between different nuclear spin species has enough time to develop before the coherence function vanishes, the product rule is no longer valid. A constructed example is shown in Fig.~\ref{fig:comparison}, where the same central spin, V and eight N nuclear spins with the same interactions as for VOPc@GNR are put in a sparse random H nuclear spin bath of number density $1/8\, \textrm{nm}^3$ which alone contributes to a $T_2$ of order $1\ms$. 
A substantial deviation of the Hahn-echo coherence function from the product of those from individual nuclear spin species occurs from $0.1\ms$ onward.

\begin{figure*}[htp]
$\begin{array}{ccc}
\includegraphics[width=0.31\linewidth]{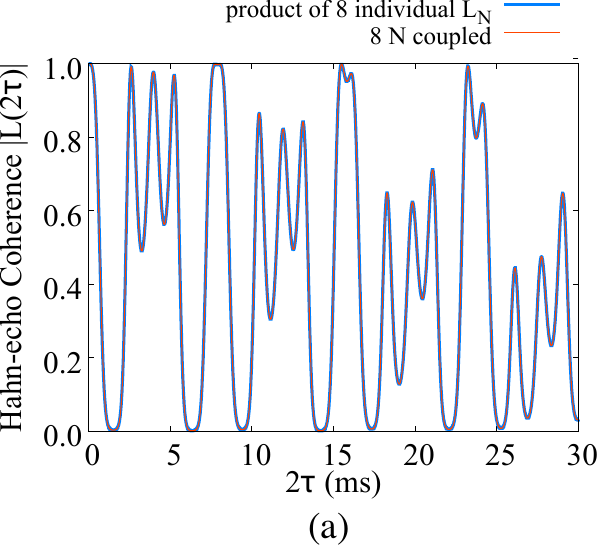}&
\includegraphics[width=0.31\linewidth]{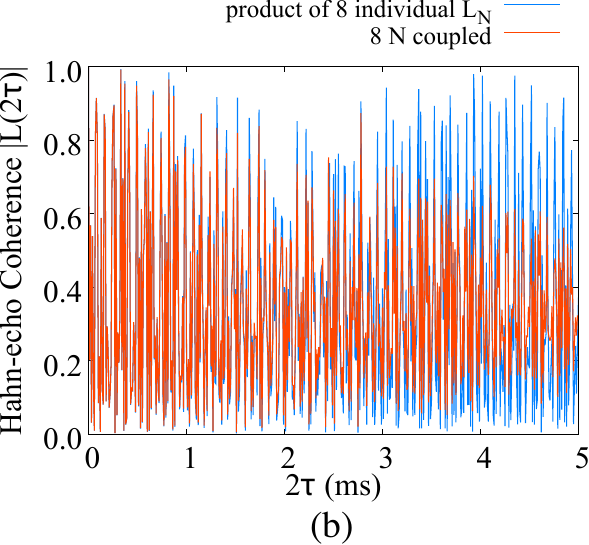}&
\includegraphics[width=0.31\linewidth]{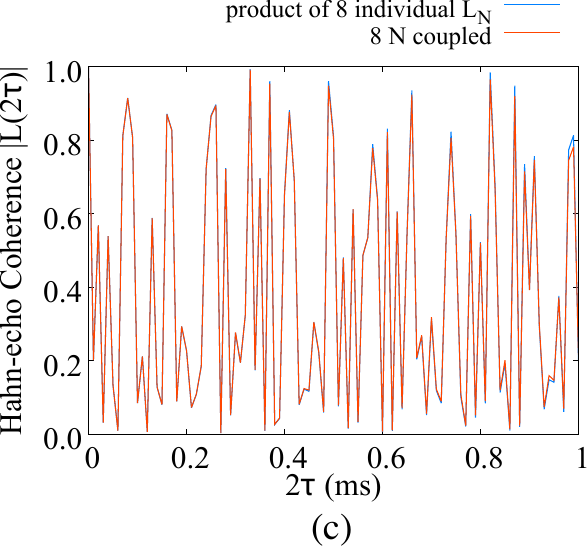}
\end{array}$
\caption{Comparison between the coherence functions due to eight N nuclear spins together and the product of coherence functions due to each individual N {(a)} over a small range of $2\tau$ comparable to $T_2$ in VOPc@GNR, {(b)} over a large range of $2\tau$ on the order of milliseconds,  {(c)} over an intermediate range of $2\tau$ where deviation between the two curves is just seen. In this example, the magnetic field of strength $0.2\T$ is along the GNR.}\label{8N_different_time_scale}
\end{figure*}

Since this product rule holds for VOPc@GNR, to understand the ESEEM in the coherence function of the full system first at small fields, all we need do is to understand the ESEEM in the coherence function when only the eight N nuclear spins are present as the bath spins, which greatly simplifies the problem. Even in this small system with only one central spin and eight N nuclear spins, a similar product rule holds at the scale of $2\tau$ smaller than one millisecond. 
In Fig.~\ref{8N_different_time_scale}(a), we show good agreement over a small range of $2\tau$ comparable to $T_2$ in VOPc@GNR between the coherence function due to eight N nuclear spins coupled together and the product of eight coherence functions calculated by including different N nuclear spin one at a time as the only bath spin in the spin Hamiltonian [{Examples for directions (ii) and (iii) are shown in Appendix~\ref{product_rules_directions_ii_and_iii}.}]. 
Deviation between the two curves is not seen until $2\tau$ is on the order of one millisecond, as shown in Fig.~\ref{8N_different_time_scale}(b) and \ref{8N_different_time_scale}(c). This means that the correlation between different N nuclear spins does not develop until a much larger time scale than $T_2$ of the system we are interested in. 
Since for the study of decoherence in the VOPc@GNR system this product rule between individual N nuclear spins is valid, now the problem can be further reduced to studying ESEEM due to individual N nuclear spins.

\begin{figure*}
\centering
\includegraphics[width=0.22\textwidth]{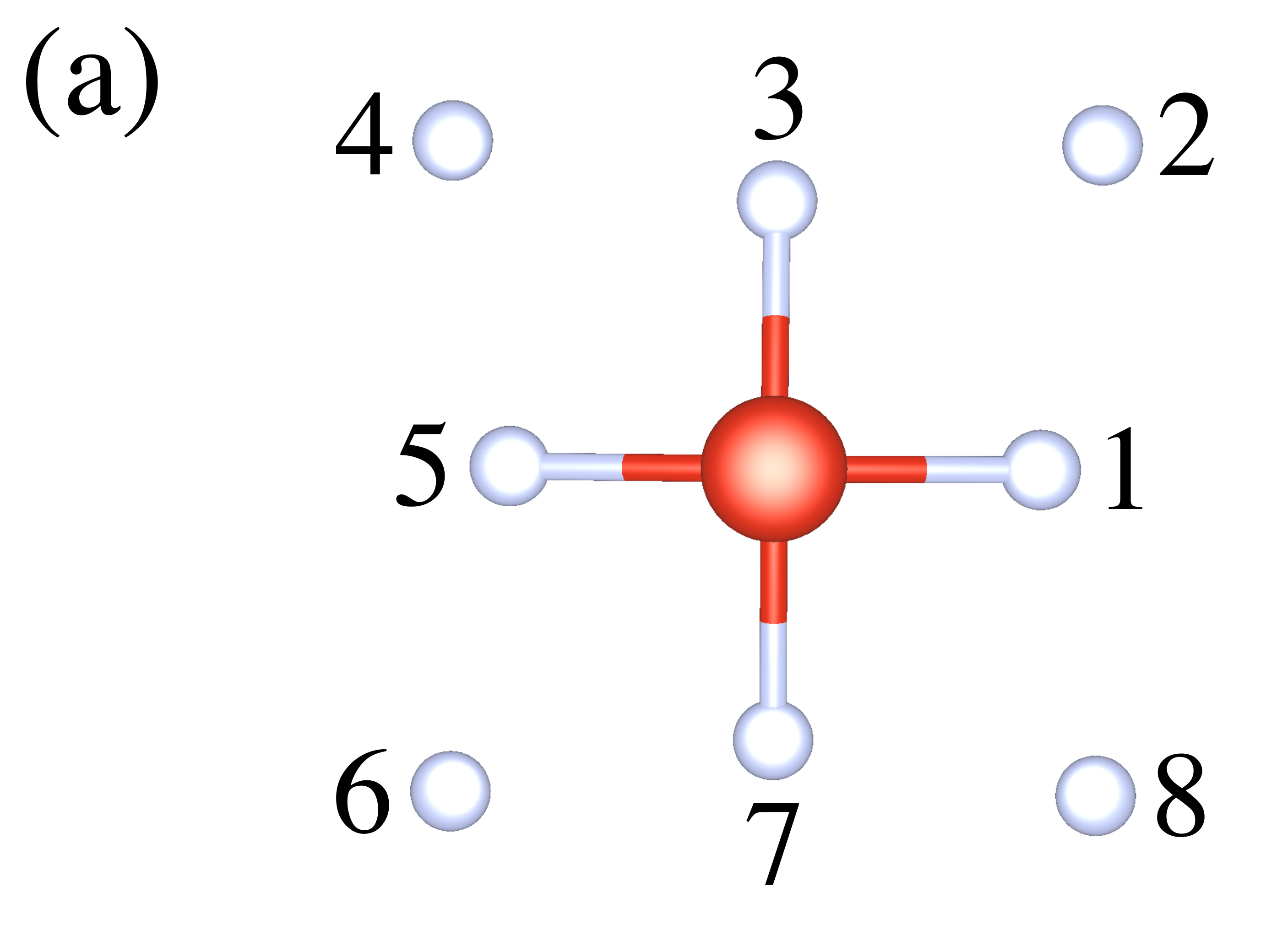}\\
$\begin{array}{cc}
\includegraphics[width=0.4\textwidth]{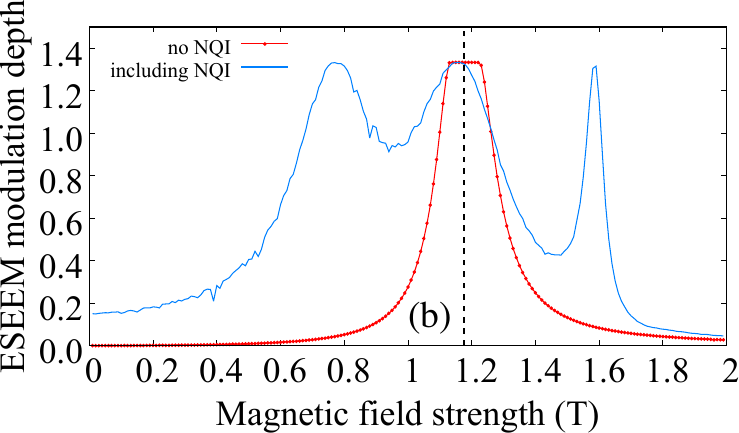}~~~~~&
\includegraphics[width=0.4\textwidth]{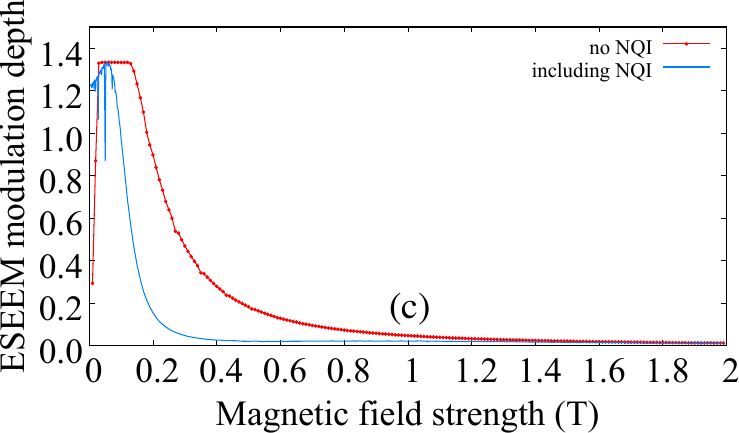}\\
\includegraphics[width=0.4\textwidth]{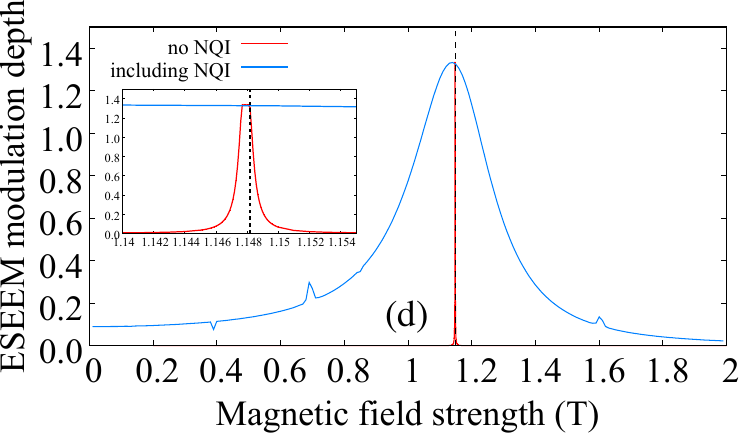}~~~~~&
\includegraphics[width=0.4\textwidth]{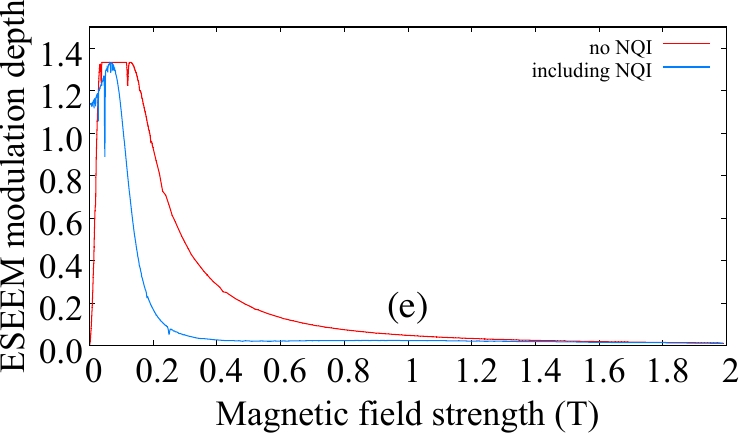}\\
\includegraphics[width=0.4\textwidth]{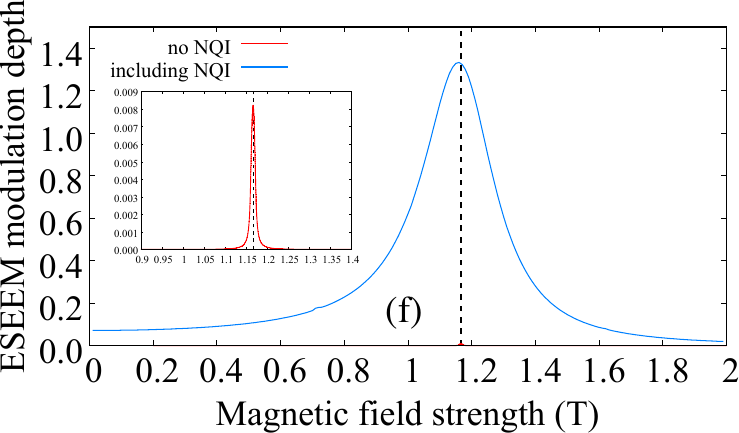}~~~~~&
\includegraphics[width=0.4\textwidth]{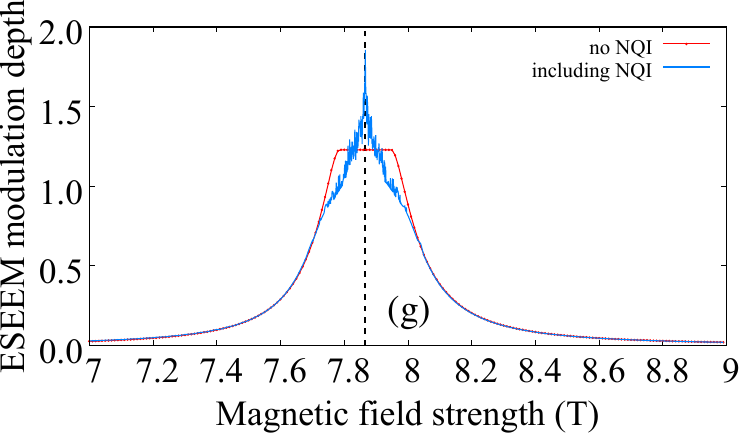}
\end{array}$
\caption{(a) Relative positions of the V and N nuclear spins viewed through the direction from the oxygen to the V nucleus. The labelling of the eight N nuclear spins N1--N8 is as shown and will be referred to in the following. The GNR (not shown) is parallel to the N5-N1 direction and below the spins in this graph.  (b) The modulation depth of ESEEM due to N1/N5 as a function of magnetic field strength. In this example, the field is along the GNR. The red curve and circles showing data sampling are for the case without the NQI. The blue curve  includes the NQI in the spin Hamiltonian. (c) Same as (b) but for N2/N4. (d) Same as (b) but for N3. Inset is a zoom-in that shows details of the red peak.  (e) Same as (b) but for N6/N8.  (f) Same as (b) but for N7. Note that the peak of the modulation depth for the case without the NQI is tiny, as shown in the inset. (g) Same as (b) but for the V nuclear spin. Vertical dashed lines in (b), (d) ,(f) and (g) indicate the value of $B_{\peak}=|{A_{zz}}|/2\gamma_{n}$.}\label{fig:modulation_depth_noQ_vs_Q}
\end{figure*}

Modulation depth of the ESEEM in the central spin coherence function when the central spin is coupled to single N nuclear spins is presented as a function of magnetic field strength in Fig.~\ref{fig:modulation_depth_noQ_vs_Q}. Here, modulation depth is defined as the maximum value of $1-L(2\tau)$ over all $2\tau$ values, the largest distance away from the full coherence of $1$ the coherence function can reach during an ESEEM oscillation. We label the eight N nuclear spins N1 to N8, as indicated in Fig.~\ref{fig:modulation_depth_noQ_vs_Q}(a), which is a view of relative positions of the V and N nuclear spins through the direction connecting the oxygen to the V nucleus. The quasi-1D GNR is parallel to the N5-N1 direction and below (not shown) N6, 7, 8. Due to symmetry, nuclei N1 and N5 cause the same (de)coherence, as well as the pair N2 and N4 and the pair N6 and N8. N3 and N7 do not cause the same central spin coherence because the GNR is closer to N7. The magnetic field dependence of the ESEEM depth due to N1/N5, N2/N4, N3, N6/N8 and N7 are shown as the red curves in Fig.~\ref{fig:modulation_depth_noQ_vs_Q}(b), (c), (d), (e) and (f). The data points sampled are marked on the curves as well. Blue curves in the same plots are results obtained by additionally including NQI in the Hamiltonian and will be discussed in Sect.~IIID. 
These results tell us that the ESEEM depth due to a single N nuclear spin has a peak on the field-strength domain centered at a position that can be different for different N positions, between 0 and 2$\T$. 
This, following the product rules, gives rise to the significant ESEEM observed in the same field range in the coherence functions due to eight coupled N nuclear spins [Fig.~\ref{fig:product_elements_1a}(b)] and from the full spin Hamiltonian [Fig.~\ref{fig:field_direction_and_strength_dependence}(a)].

A two-spin model study 
to understand why the nuclear spin modulation depth for a single N peaks at a certain field strengths is presented in Sect. III(C). The ESEEM due to the V nuclear spin is also addressed there.

\subsection{Simple model study of ESEEM \label{two-spin_model}}

We consider a simple two-spin model from which we can obtain a closed form expression for the oscillations in the Hahn-echo coherence function.
The system consists of one electron spin-${1}/{2}$  and one N nuclear spin-$1$. The spin Hamiltonian, not including NQI, is
\beq
\hat{H}=-\gamma_{e}\bm{B}\cdot\hat{\bm{S}}-\gamma_{N}\bm{B}\cdot\hat{\bm{I}}+\hat{\bm{S}}\cdot\mathbf{A}\cdot\hat{\bm{I}}, \label{eq:two_spin_full_H}
\eeq
following Eqs.~(\ref{spin_Hamiltonian})--(\ref{bath}).
By simulating the electron spin Hahn-echo coherence function for this Hamiltonian with first-principles inputs for the hyperfine interaction tensor $\mathbf{A}$, we obtained results on the ESEEM due to individual N nuclear spins reported in Sect.~III(B).
Without loss of generality, we will set the $z$ direction in the model along the magnetic field. 
In order to obtain a closed form expression for the coherence function, a secular approximation is applied where only the terms in the hyperfine interaction containing $\hat{S}_{z}$, \textit{i.e.} $A_{zi}\hat{S}_{z}\hat{I}_{i}$ ($i=x$, $y$, $z$), are kept and other hyperfine terms are dropped. The secular approximation requires the electron spin Zeeman splitting to be much larger than all other terms in the spin Hamiltonian, which is the case for the spin interactions in VOPc-GNR at all the field strengths we consider.
After a rotation of the $x$ and $y$ axes about $z$, $A_{zy}$ can be reduced to zero, further simplifying $\hat{H}$.
The Hamiltonian is now 
\beq
\hat{H}=\omega_{e}\hat{S}_{z}+\omega_{N}\hat{I}_{z}+A_{zx}\hat{S}_{z}\hat{I}_{x}+A_{zz}\hat{S}_{z}\hat{I}_{z},
\eeq
where $\omega_{e}=-\gamma_{e}B$ and $\omega_{N}=-\gamma_{N}B$.
The third and fourth terms represent the pseodosecular and secular part of the hyperfine interaction, respectively.

To simulate the Hahn-echo experiment, the initial density matrix following the first, ${\pi}/{2}$ pulse in the Hahn-echo pulse sequence is again described by Eqs.~(\ref{eq:initial_total_dm})--(\ref{initial_B_state}). Its operator form can be written as
\beq
\hat{\rho}(0^{+}) = 
(\hat{S}_{x}+\textstyle{\frac{1}{2}}\hat{S}_{0})\otimes
(\textstyle{\frac{1}{3}}\hat{I}_{0}),
\eeq
where $\hat{S}_{0}$ and $\hat{I}_{0}$ are identity operators in the state spaces of the electron and the nuclear spin, respectively.
The goal is to find the Hahn-echo coherence function $L(2\tau)$ of the electron spin from the system 
density matrix at $t=2\tau$, with a $\pi$-pulse applied to the electron spin at
$t=\tau$. Following Eq.~(\ref{coherence_function}),
\beq
L(2\tau)=2\Tr[\hat{\rho}(2\tau)(\hat{S}_{x0}-i\hat{S}_{y0})] , 
\label{coherence} 
\eeq
\beq
\hat{\rho}(2\tau)=e^{-i\hat{H}\tau}e^{-i\pi\hat{S}_{x0}}e^{-i\hat{H}\tau}\hat{\rho}(0^{+})e^{i\hat{H}\tau}e^{i\pi\hat{S}_{x0}} e^{i\hat{H}\tau},
\eeq
where $\hat{S}_{x0}=\hat{S}_{x}\hat{I}_{0}$ and $\hat{S}_{y0}=\hat{S}_{y}\hat{I}_{0}$. This model was first considered by W. B. Mims in Ref.~[{\onlinecite{mims1972envelope}}], in which he obtained an analytical expression for $L(2\tau)$.
We have reproduced the solution and apply it to study the ESEEM depth from N nuclear spins in VOPc@GNR.
The closed form expression for $L(2\tau)$ is rather long, and we present it in Eqs.~(\ref{eq:two_spin_model_L_expression}) and (\ref{ESEEM_coefficients}) in Appendix \ref{model}.
The ESEEM of a single frequency is described by a cosine term in Eq.~(\ref{eq:two_spin_model_L_expression}), and its modulation depth is just the absolute value of the coefficient in front of it.
The modulation depths of the ESEEM of all frequencies share a common factor $C$,
\beq
C=\frac{1}{3\bigl[A_{zx}^{4}+2(A_{zz}^{2}+4B^{2}\gamma_{N}^{2})A_{zx}^{2}+(A_{zz}^{2}-4B^{2}\gamma_{N}^{2})\strut^{2}\bigr]^{2}},\label{eq:common_factor}
\eeq
with the remaining factors in the coefficients for each frequency expressed as $a$, $b$, $c$, $\dots$, $l$ in Eq.~(\ref{ESEEM_coefficients}).

\begin{figure}[htp]
\includegraphics[width=0.95\linewidth]{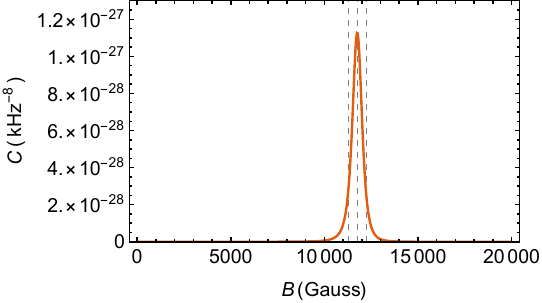}
\caption{The common factor in modulation depths of all ESEEM frequencies in the two-spin model, $C$, as a function of the magnetic field strength $B$, for the N1 nuclear spin when the field is along the GNR.  Three dashed lines indicate the field strengths $B_{\peak}$ and $B_{\peak} \pm |A_{zx}|/2\gamma_{N}$. }\label{fig:N1_C_B_along_GNR}
\end{figure}

According to the behavior of $C$, there are two different scenarios relevant to N nuclear spins in VOPc@GNR: 
The first is when $|A_{zz}| \gg |A_{zx}|$, which is the case for the nearest N nuclear spins to the central spin, \textit{i.e.} N1, N3, N5 and N7. For these spins, the isotropic Fermi contact part of the hyperfine interaction [Eq.~(\ref{Fermi_contact_hyperfine})] is much larger than the anisotropic dipolar part [Eq.~(\ref{dipolar_hyperfine})], and therefore $|A_{zz}| \gg |A_{zx}|$ is valid whatever the field direction and correspondingly the $z$-axis in the model. In this scenario, $C(B)$ as a function of $B$ strongly peaks at the value of $B$ which satisfies $|{A_{zz}}|=2B{\gamma_{N}}$, which we label $B_{\peak}$. This is because when this condition is satisfied the second term in the square bracket in Eq.~(\ref{eq:common_factor}) dominates over others, leading to the maximum of $C(B)$, $C_{max} \approx (1/48)A_{zz}^{-4}A_{zx}^{-4}$, while when $|2B{\gamma_{N}}-|{A_{zz}}||$ is on the order of $|{A_{zz}}|$, $C \sim A_{zz}^{-8}$, and $C$ is even smaller if $2B{\gamma_{N}}$ further deviates from $|{A_{zz}}|$. The width of the peak is measured by $|{A_{zx}}|/2\gamma_{N}$, as $C(B_{\peak} \pm |A_{zx}|/2\gamma_{N}) \approx (1/192)A_{zz}^{-4}A_{zx}^{-4}=(1/4)C_{max}$. As an example, for N1 when the field is along the GNR, the corresponding hyperfine interactions, obtained from DFT, are $|{A_{zz}}|=7242\kHz$ and $|{A_{zx}}|=286\kHz$. 
The corresponding $C(B)$ is shown in Fig.~\ref{fig:N1_C_B_along_GNR} where the values of $B_{\peak}$ and $B_{\peak} \pm |A_{zx}|/2\gamma_{N}$ are also labelled by the vertical dashed lines. 

\begin{figure}[htp]
\includegraphics[width=0.95\linewidth]{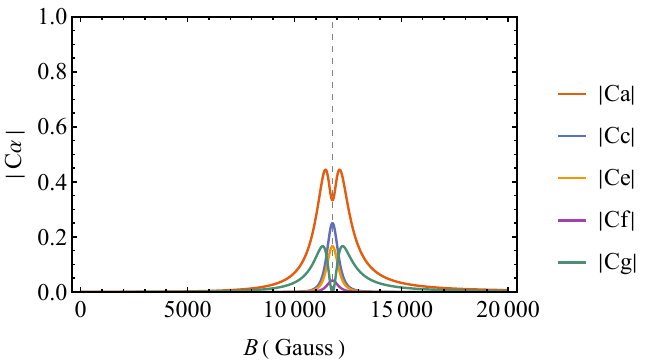}
\caption{Independent values of the modulation depths $|C\alpha|$ [Eq.~(\ref{ESEEM_coefficients})] of single ESEEM frequencies, for the N1 nuclear spin when the field is along the GNR. The dashed line indicates the field strengths $B_{\peak}$.
}\label{fig:N1_Cacefg_B_along_GNR}
\end{figure}

Since the remaining factor in the modulation depth for each ESEEM frequency, $a$, $b$, $c$, $\dots$,  $l$ are functions of $B$ which only show mild variation, the modulation depth $|C\alpha|$ ($\alpha=a$, $b,\dots$, $l$) also have significant strength only in a narrow range of magnetic field around $B_{\peak}$. For the example of N1 above, $|C\alpha|$ are plotted in Fig.~\ref{fig:N1_Cacefg_B_along_GNR}. Since the total ESEEM due to a single N nuclear spin is just the sum of the ESEEM of all frequencies [Eq.~(\ref{eq:two_spin_model_L_expression})], it has modulation depth reaching its maximum also at the field strength of $B_{\peak}$, which explains the location of the modulation depth peaks of N1/N5, N3 and N7 in the red curves in Fig.~\ref{fig:modulation_depth_noQ_vs_Q}(b), (d) and (f), respectively. The value of $B_{\peak}$ is indicated in Fig.~\ref{fig:modulation_depth_noQ_vs_Q}(b), (d) and (f) as the vertical dashed line.

\begin{figure}[htp]
\includegraphics[width=0.95\linewidth]{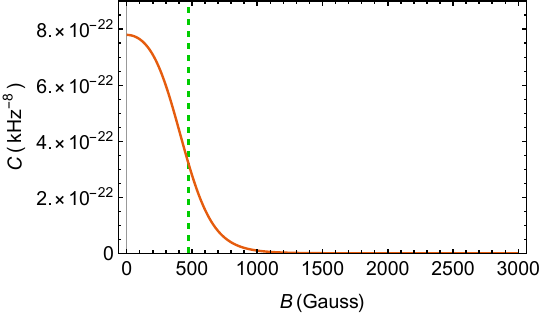}
\caption{The common factor in modulation depths of all ESEEM frequencies in the two-spin model, $C$, as a function of the magnetic field strength $B$, for the N2 nuclear spin when the field is along the GNR.  The green dashed line indicates the field strengths $|A_{zx}|/2\gamma_{N}$. }\label{fig:N2_C_B_along_GNR}
\end{figure}

The second scenario is when $|A_{zz}|$ and $|A_{zx}|$ are of the same order, which is the case for the N nuclear spins farther away from the central spin, N2, N4, N6 and N8, and for the three directions of field, which defines $z$ in the model, described in Sect. III(B). $C(B)$ in this scenario has relatively significant value in the range of field strength starting from zero to a value of the order of $|A_{zx}|/\gamma_{N}$, compared to fields beyond the range. If $|A_{zz}| \le |A_{zx}|$, $C(B)$ has its maximum at zero field and is a monotonically decreasing function of $B$. For the special case of $|A_{zz}|=|A_{zx}|$, $C(|A_{zx}|/2\gamma_{N})=(16/25) \, C(B=0)$. As an example, for N2 when the field is along the GNR, the corresponding hyperfine interactions, obtained from DFT, are $|{A_{zz}}|=243\kHz$ and $|{A_{zx}}|=291\kHz$. The corresponding $C(B)$ is plotted in Fig.~\ref{fig:N2_C_B_along_GNR} where the value of $|A_{zx}|/2\gamma_{N}$ is labelled by the green dashed line.

\begin{figure}[htp]
\includegraphics[width=0.95\linewidth]{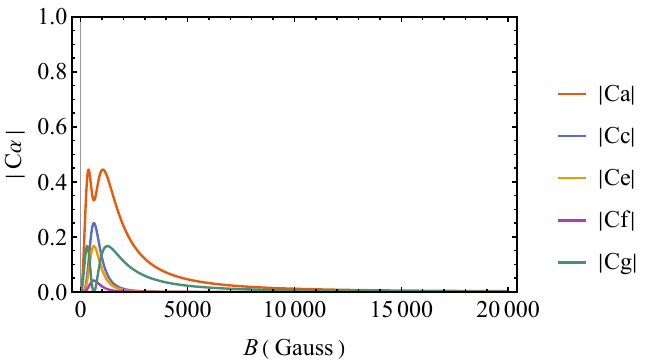}
\caption{Independent values of the modulation depths $|C\alpha|$ [Eq.~(\ref{ESEEM_coefficients})] of single ESEEM frequencies, for the N2 nuclear spin when the field is along the GNR.
}\label{fig:N2_Cacefg_B_along_GNR}
\end{figure}

Similar to the first scenario, constrained by the range of fields where $C(B)$ is prominent, the modulation depth $|C\alpha|$ ($\alpha=a,b,...,l$) can have significant value only in a narrow range of low magnetic field. When approaching zero field, $|C\alpha|$ vanish due to a, b, ..., l [Eq.~(\ref{ESEEM_coefficients})] being polynomials of $B$ without a constant term.
For the example of N2 above, $|C\alpha|$ are plotted in Fig.~\ref{fig:N2_Cacefg_B_along_GNR}.
Similar to the first scenario, this explains why modulation depth becomes large at low fields for N2/N4, N6/N8 but vanish at zero field [red curves in Fig.~\ref{fig:modulation_depth_noQ_vs_Q}(c), (e)].

Now we have an understanding of why the modulation depths of ESEEM due to individual N nuclear spins become large near certain field strengths within the range 0--2$\T$. For the V nuclear spin, an analytical study of modulation depth in a $S={1}/{2}$, $I={7}/{2}$ model is much more complicated than the $S={1}/{2}$, $I=1$ model above for N, and a closed form expression may even be impossible. Here we simply state that for the V nuclear spin in VOPc@GNR and the three directions we consider in Sect.~III(B), $|A_{zz}| \gg |A_{zx}|$ from DFT and the qualitative result of the first scenario in the $S={1}/{2}$, $I=1$ model above, \textit{i.e.} ESEEM depth caused by the nuclear spin reaches maximum at $B_{\peak}=|{A_{zz}}|/2\gamma_{n}$, $\gamma_{n}$ being the nuclear gyromagnetic ratio, is also valid. The central spin ESEEM depth in VOPc@GNR due to the V nuclear spin when the field is along the GNR is plotted as a function of field strength by the red curve in Fig.~\ref{fig:modulation_depth_noQ_vs_Q}(g). The value of $B_{\peak}=|{A_{zz}}|/2\gamma_\textrm{V}$ is labelled by the vertical dashed line.



\subsection{Effect of nuclear quadrupole interaction \label{SubSect:NQI_effect}}

In this section, we describe the effect of the nuclear quadrupole interaction (NQI)  on the Hahn-echo coherence functions. The spin Hamiltonian now includes the NQI term $\sum_{i}\hat{\bm{I}}_{i}\cdot\mathbf{P}_{i}\cdot\hat{\bm{I}}_{i}$ [cf.~eq.~(\ref{bath})], where the quadrupole interaction tensor $\mathbf{P}_{i}$ of nuclear spin $i$ is proportional to the EFG tensor at its position, 
\beq
\mathbf{P}_{i}=\frac{eQ_{i}}{2I_{i}(2I_{i}-1)h}\mathbf{V}_{i}, 
\label{quadrupole} 
\eeq
where $e$ is the elementary charge, $Q_{i}$ the nuclear electric quadrupole moment, $I_{i}$ the nuclear spin quantum number, and $h$ the Planck constant. The EFG tensor $\mathbf{V}_{i}$, the second order derivative of the electrostatic potential at the position of nuclear spin $i$ due to all charges external to the nucleus 
with components $(\mathbf{V}_{i})_{\alpha\beta}={\partial^{2}V(\bm{R}_{i})}/{\partial\alpha \, \partial\beta}$, $\alpha$, $\beta=x$, $y$, $z$, is obtained from DFT calculations.


\begin{figure*}[htp]
$\begin{array}{ccc}
\includegraphics[width=0.31\linewidth]{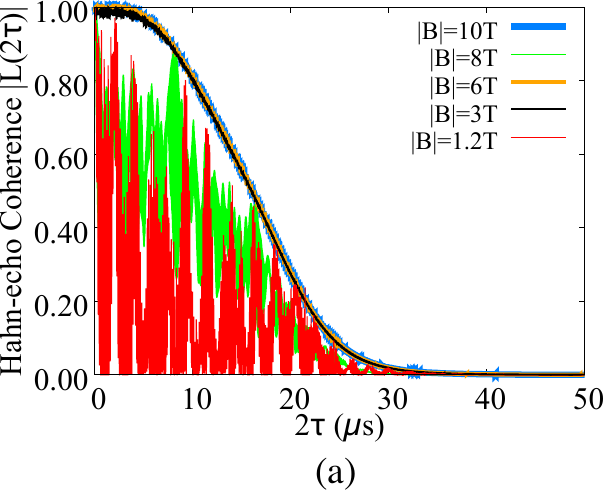}&
\includegraphics[width=0.31\linewidth]{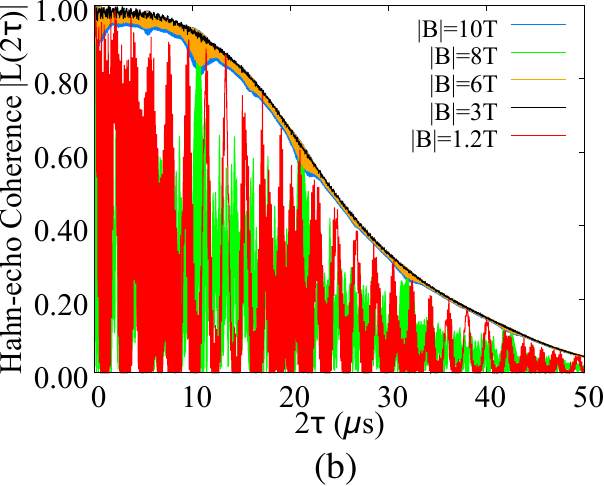}&
\includegraphics[width=0.31\linewidth]{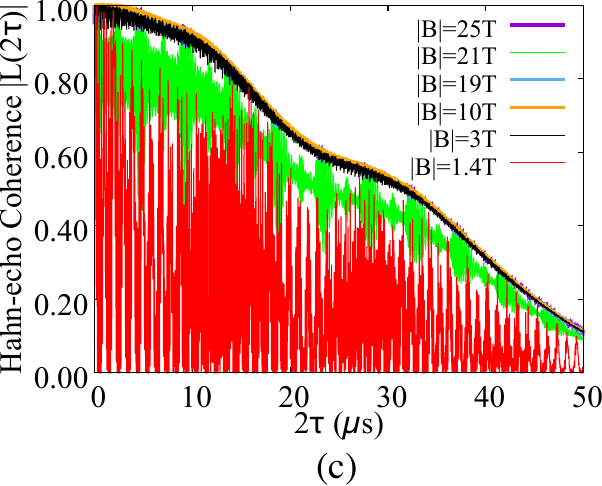}
\end{array}$
\caption{Dependence of the Hahn echo coherence functions on magnetic field strength for field directions as in Fig.~\ref{fig:field_direction_and_strength_dependence} using the spin Hamiltonian including the NQI: {(a)} magnetic field in the direction along the GNR; {(b)} magnetic field in direction (ii) described in Sect.~IIIB; {(c)} magnetic field in the direction parallel to the V-O bond. }\label{fig:field_direction_and_strength_dependence_with_quadrupole}
\end{figure*}

\begin{figure*}[htp]
$\begin{array}{cc}
\includegraphics[width=0.4\linewidth]{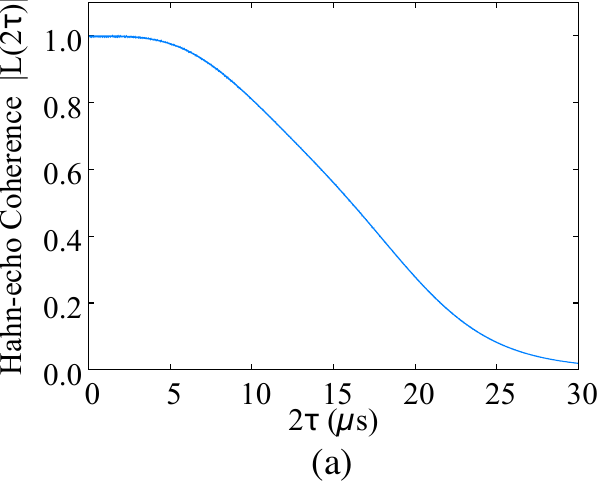}&
\includegraphics[width=0.4\linewidth]{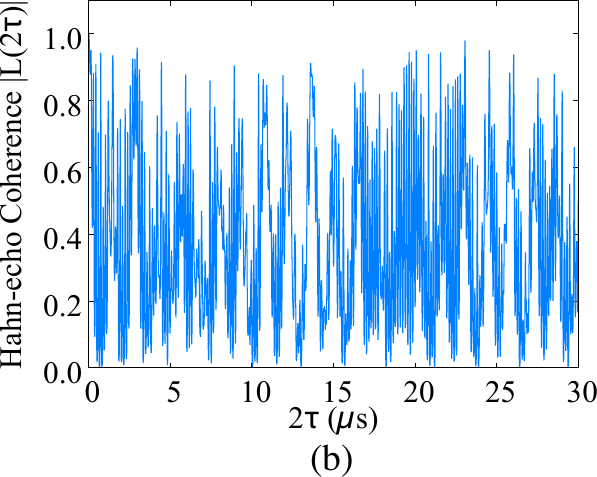}\\
\includegraphics[width=0.4\linewidth]{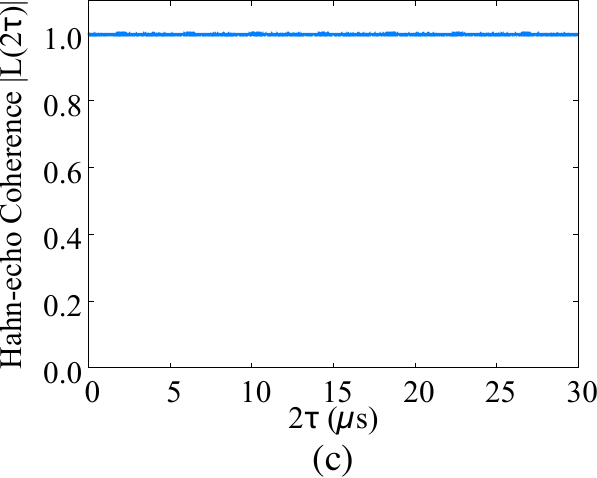}&
\includegraphics[width=0.4\linewidth]{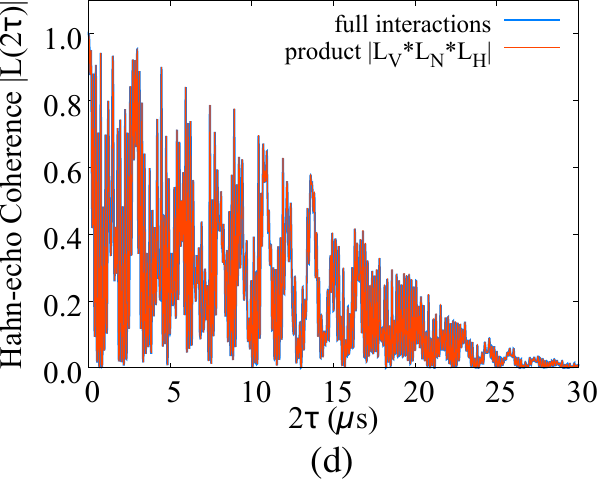}
\end{array}$
\caption{Product rule of the Hahn-echo coherence function among different elements in the VOPc@GNR system using the spin Hamiltonian including the NQI. A magnetic field of strength $1\T$ is applied in the direction along the GNR as an example. The coherence functions are calculated from the central spin coupling to {(a)} only the H nuclei present as bath spins,  {(b)} only the N nuclei present as bath spins, and {(c)} only the V nucleus present as the bath spin. In {(d)}, the coherence function due to all elements present in the spin Hamiltonian (blue) is seen to closely follow the product (red) of (a), (b) and (c). }\label{fig:product_elements_1T_with_quad}
\end{figure*}

\begin{figure*}[htp]
$\begin{array}{cc}
\includegraphics[width=0.4\linewidth]{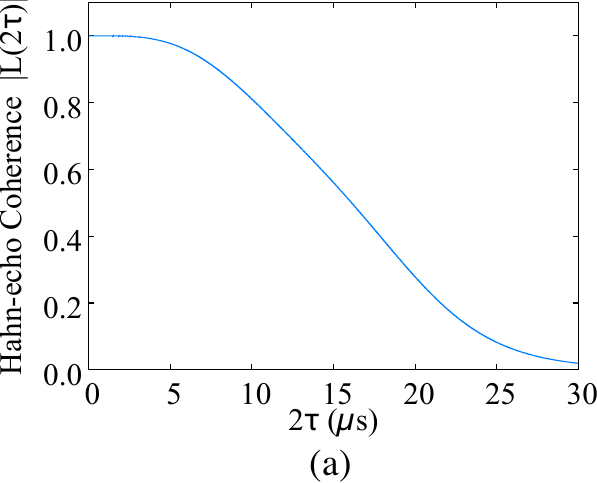}&
\includegraphics[width=0.4\linewidth]{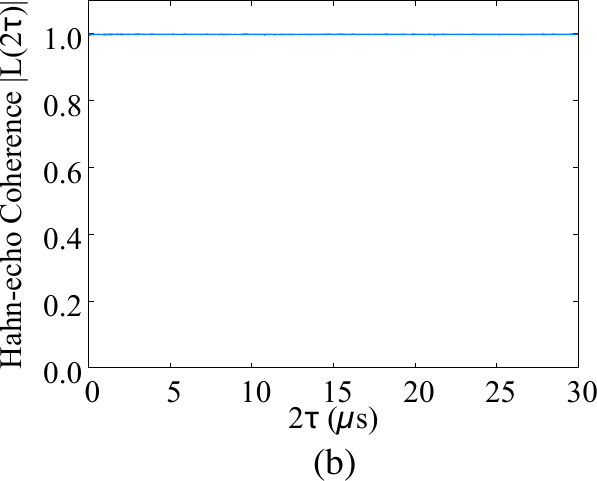}\\
\includegraphics[width=0.4\linewidth]{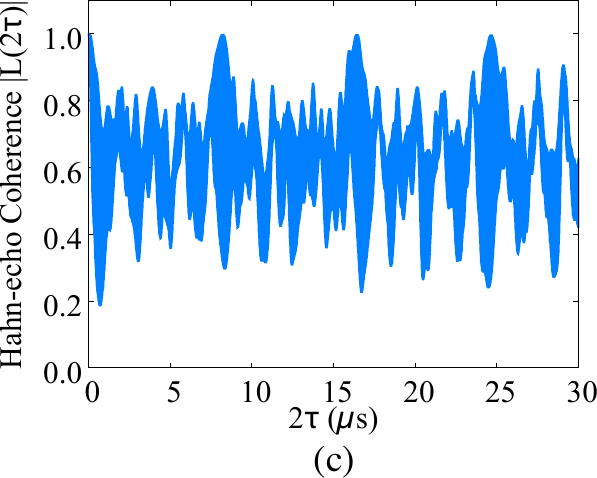}&
\includegraphics[width=0.4\linewidth]{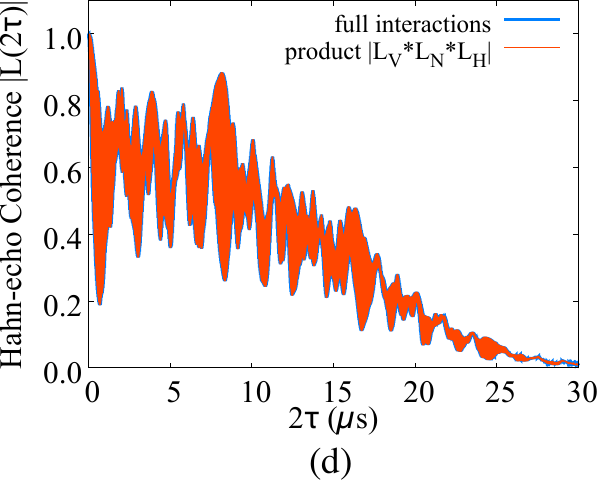}
\end{array}$
\caption{Product rule of the Hahn-echo coherence function among different elements in the VOPc@GNR system using the spin Hamiltonian including the NQI. A magnetic field of strength $8\T$ is applied in the direction along the GNR as an example. The coherence functions are calculated from the central spin coupling to {(a)} only the H nuclei present as bath spins,  {(b)} only the N nuclei present as bath spins, and {(c)} only the V nucleus present as the bath spin. In {(d)}, the coherence function due to all elements present in the spin Hamiltonian (blue) is seen to closely follow the product (red) of (a), (b) and (c). }\label{fig:product_elements_8T_with_quad}
\end{figure*}

\begin{figure}[htp]
\includegraphics[width=0.8\linewidth]{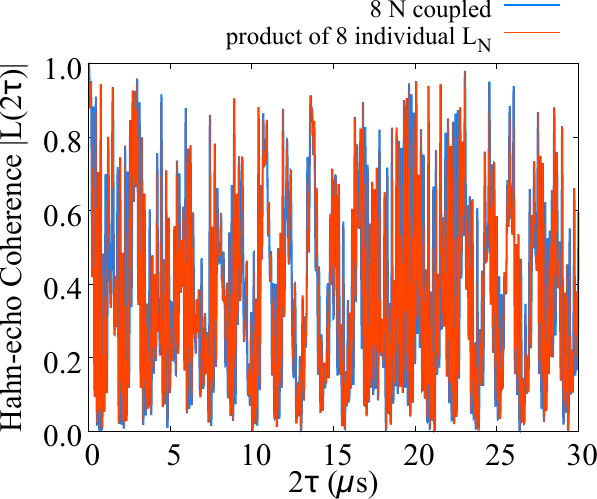}
\caption{Similar to Fig.~\ref{8N_different_time_scale}(a), an agreement is found between the coherence functions due to eight N nuclear spins coupled together (blue) and the product of coherence functions due to each individual N (red). Here the NQI is included. A magnetic field of $1\T$ in this example is applied along the GNR.}\label{fig:8N_with_quadrupole}
\end{figure}

Our simulations show that the inclusion of the NQI for V and N nuclear spins does not alter the following results: 
(1) The general ranges of magnetic field strength where large ESEEM appears are not significantly changed. For magnetic fields along the GNR and along direction (ii) as in Sect.~III(B), significant ESEEM is still present both at the relatively small fields of 0--$2\T$ and large fields of 7--$9\T$, while it is suppressed in the intermediate field range of 3--$6\T$ [Fig.~\ref{fig:field_direction_and_strength_dependence_with_quadrupole}(a) and \ref{fig:field_direction_and_strength_dependence_with_quadrupole}(b)]. For the case of the field parallel to the V-O bond, ESEEM is still present only in the field ranges of 0--$2\T$ and 20--$24\T$ [Fig.~\ref{fig:field_direction_and_strength_dependence_with_quadrupole}(c)]. 
(2) The product rule of Hahn-echo coherence functions between different nuclear spin species, as in Figs.~\ref{fig:product_elements_1a} and \ref{fig:product_elements_1b}, still holds (Figs.~\ref{fig:product_elements_1T_with_quad} and \ref{fig:product_elements_8T_with_quad}). Therefore the coherence time $T_2$ of the envelopes of the coherence functions remain unchanged, since the envelopes are contributed by H nuclear spins, which are not affected by the NQI. 
(3) The product rule for the Hahn-echo coherence function between different nitrogen nuclear spins still holds (Fig.~\ref{fig:8N_with_quadrupole}); therefore the variation of the modulation depth of the ESEEM due to all nitrogen nuclear spins can again be understood by those due to individual spins (blue curves in Fig.~\ref{fig:modulation_depth_noQ_vs_Q}). 
(4)  For the ESEEM due to individual nuclear spins, the modulation depth (blue curves in Fig.~\ref{fig:modulation_depth_noQ_vs_Q}) still reaches a maximum at approximately the same magnetic field as the case without the NQI (red curves in Fig.~\ref{fig:modulation_depth_noQ_vs_Q}).

\begin{figure}
\centering
$\begin{array}{c}
\includegraphics[width=0.45\textwidth]{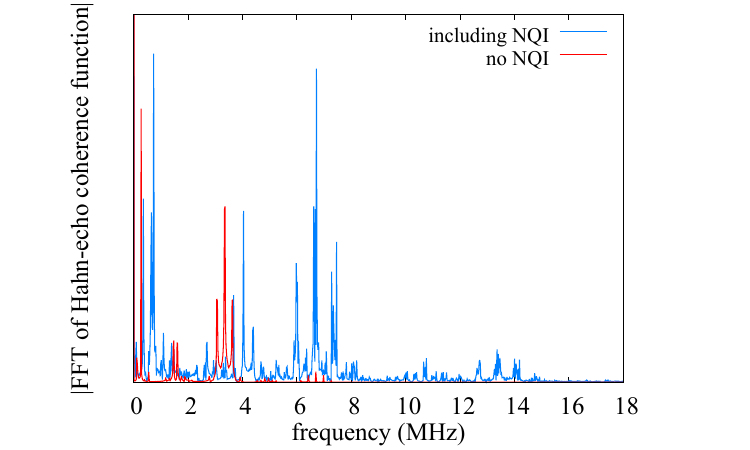}\\
\includegraphics[width=0.45\textwidth]{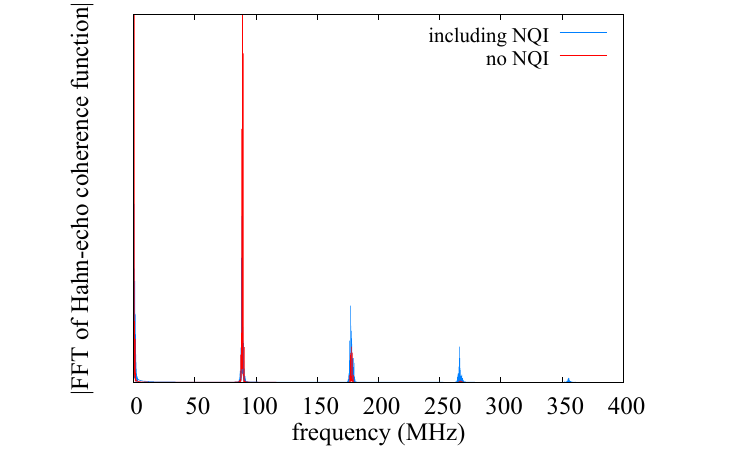}\\
\includegraphics[width=0.45\textwidth]{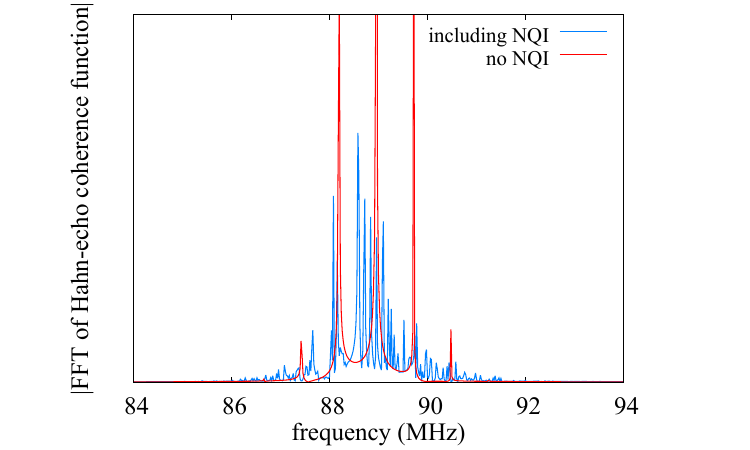}
\end{array}$

\caption{Comparison between the magnitude of the Fourier transform of the ESEEM due to the spin Hamiltonian including (blue curves) or not including (red curves) the NQI. Here as an example the magnetic field is along the GNR direction. 
(Top) Field strength $B=1\T$. The blue curve is the magnitude of the Fourier transform of Fig.~\ref{fig:product_elements_1T_with_quad}(b) rather than Fig.~\ref{fig:product_elements_1T_with_quad}(d) in order to better resolve frequencies related to nitrogen nuclear spins, which are responsible for the ESEEM at this field strength. The red curve is the same quantity in the absence of NQI. 
(Middle) Field strength $B=8\T$. Blue and red curves are Fourier transforms of Fig.~\ref{fig:product_elements_8T_with_quad}(c) and \ref{fig:product_elements_1b}(c), respectively. Peak structures appear with a period of around $89\MHz$. (Bottom) A zoom-in of the peak structure in the range of 84--$94\MHz$ in the middle panel showing a change in the detailed satellite structure of the frequency peaks once the NQI is included.}\label{fig:ESEEF_fft_noQ_vs_Q}

\end{figure}

Inclusion of the NQI introduces two major changes. One is a change in ESEEM frequencies, as shown in the magnitude of the ESEEM Fourier transforms in Fig.~\ref{fig:ESEEF_fft_noQ_vs_Q}(Top) where we compare the frequencies of the ESEEM in Fig.~\ref{fig:product_elements_1T_with_quad}(b) which is due to N nuclear spins including the NQI with that without the NQI at the same field. The same comparison at $8\T$ between the frequencies of the ESEEM due to the V nuclear spin with and without NQI, as in Fig.~\ref{fig:product_elements_1b}(c) and \ref{fig:product_elements_8T_with_quad}(c), is shown in Fig.~\ref{fig:ESEEF_fft_noQ_vs_Q}(Middle). A zoom-in of the first positive-frequency peak structure in Fig.~\ref{fig:ESEEF_fft_noQ_vs_Q}(Middle) is displayed in \ref{fig:ESEEF_fft_noQ_vs_Q}(Bottom), showing a change of the detailed satellite structure due to the NQI. The positions of the peak structures in Fig.~\ref{fig:ESEEF_fft_noQ_vs_Q}(Middle) are periodic with a frequency spacing of around $89\MHz$, which approximately corresponds to the Zeeman splitting of the V nuclear spin at this field, $89.7\MHz$. 
The second major change is that for the ESEEM due to individual nuclear spins, although the modulation depth still reaches maximum at approximately the same magnetic field as the case without NQI,
NQI can change the height and width of the peaks in the modulation depth as a function of magnetic field [Fig.~\ref{fig:modulation_depth_noQ_vs_Q}(c)--(g)] and even introduces additional peaks on both sides [Fig.~\ref{fig:modulation_depth_noQ_vs_Q}(b)]. When NQI is included, The modulation depth can also be nonzero when the field approaches zero, in contrast to the zero modulation depth in the same limit when NQI is absent [Fig.~\ref{fig:modulation_depth_noQ_vs_Q}(c)--(f)].

\section{Conclusion}\label{sec:Conclusion}
In this work, we have performed first-principles calculations of central spin decoherence in a nuclear spin bath for the system of VOPc@GNR. Low energy isomeric atomic configurations of the ground state as well as the corresponding electronic structures are calculated by DFT, which shows that after the integration onto the GNR, the molecular electronic spin of the VOPc molecule remains spin-${1}/{2}$ and still contributed by an unpaired $d_{xy}$ electron on the V atom. 

In the study of spin decoherence, using the CCE method with a spin Hamiltonian in which the hyperfine and NQI tensors are calculated from DFT, the time evolution of the coherence function as the off-diagonal element of the central spin RDM is examined in simulation of Hahn-echo experiments.
The central spin decoherence is found to be mainly contributed by nuclear spins within a distance of around $26\ang$ to the central spin. A comparison between the spin decoherence in this VOPc@GNR system to that in a single VOPc molecule also shows a strong decrease of $T_2$ due to the protons on the GNR. 
Three mutually perpendicular directions for the magnetic field are considered and an anisotropy in $T_2$ is observed, with the value of $T_2$ when the field is along the V-O bond almost twice as large as that when the field is along the GNR. 
Large ESEEM appears in certain ranges of magnetic field while being suppressed outside these ranges. 
A detailed investigation of the coherence functions due to individual nuclear spin species reveals a product rule, that spin coherence function due to the full spin Hamiltonian
agrees with the product of central spin coherence functions due to individual nuclear spin species.
This product rule, valid at small time scales, allows us to identify that the envelope of the coherence function, and therefore $T_2$ in VOPc@GNR, is contributed only by the H nuclear spins and that large ESEEM at relatively small fields is due to N nuclear spins while that at large fields is due to V nuclear spins. 
In the study of the ESEEM due to N nuclei, a similar product rule is found valid for the collective result for all nuclei and product of results from individual spins, reducing the problem to computing the ESEEM due to individual N nuclei.
By investigating the closed-form expressions of the Hahn-echo coherence function and the ESEEM depth in an $S={1}/{2}$, $I=1$ two-spin model, we find a relation between, on one side, the range/value of the field strength where ESEEM depth due to an individual N nuclear spin becomes significant/reaches its maximum and, on the other side, the secular  $|A_{zz}|$ and the pseudosecular $|A_{zx}|$ parts of the hyperfine interaction.
This relation explains why ESEEM due to N nuclei is present at relatively low fields. 
The qualitative result of the scenario for $|A_{zz}| \gg |A_{zx}|$ in the model, which states that the modulation depth reaches maximum at a field when the secular hyperfine interaction is equal to double the nuclear Zeeman splitting, is found to also apply to 
the V nuclear spin.
Finally, we include the nuclear quadrupole interaction calculated for the bare VOPc@GNR structure, which is not negligible for N and V nuclei, in the spin Hamiltonian and analyzed its effects. Simulation shows that while the NQI does not change the product rules, the decoherence time $T_2$ or the value of the magnetic field where modulation depth of ESEEM due to an individual nuclear spin reaches maximum,
it modifies frequencies of ESEEM oscillations, can change the width of the peak in the modulation depth as a function of magnetic field, can introduces additional peaks around the central one and can make the modulation depth nonzero in the limit of zero field. 
We thus have identified the applicability and limitation of the spin Hamiltonian without NQI that represents an incomplete description of physics in VOPc@GNR systems.

Our work provides information on the atomic configuration and the electronic structure of VOPc@GNR and can guide further experiments on this system in identifying the optimal magnetic field direction and strengths where the coherence time $T_2$ due to central electronic spin coupling to nuclear bath spins is maximized and the ESEEM is suppressed. The coherence function product rules show that this $T_2$ is constrained only by H nuclear spins, even if V and N nuclei are closer to the electron spin and have hyperfine interactions orders of magnitude larger, and confirm that the major source of spin decoherence in hydrogen-rich magnetic molecular systems is the H nuclear spins.\cite{zecevic1998dephasing,canarie2020quantitative,chen2020decoherence}
The finding of a relation between the secular/pseudosecular hyperfine interaction and the ESEEM depth provides insight for future design of magnetic molecular spin qubits to reduce ESEEM effects. 
In general, this work shows the capability of combining the DFT and the CCE methods in predicting all the details of the spin decoherence in molecular spin qubit architecture due to central electronic spin-nuclear spin coupling and provides useful insights for future designs of molecular spin-qubit architectures.

{\bf Acknowledgements.}  The authors are grateful for useful conversations with Silas Hoffman, Shuanglong Liu, Haechan Park, Steve Hill. This work is supported by the U.S. Department of Energy, Office of Science, Basic Energy Sciences under Award No. DE-SC0022089. Computations were done using the utilities of the National Energy Research Scientific Computing Center and University of Florida Research Computing.

\clearpage

\bibliography{main}

\null
\vskip 0.5in

\appendix 

\begin{figure*}[ht]
\centering
$\begin{array}{cc}
\includegraphics[width=0.45\textwidth]{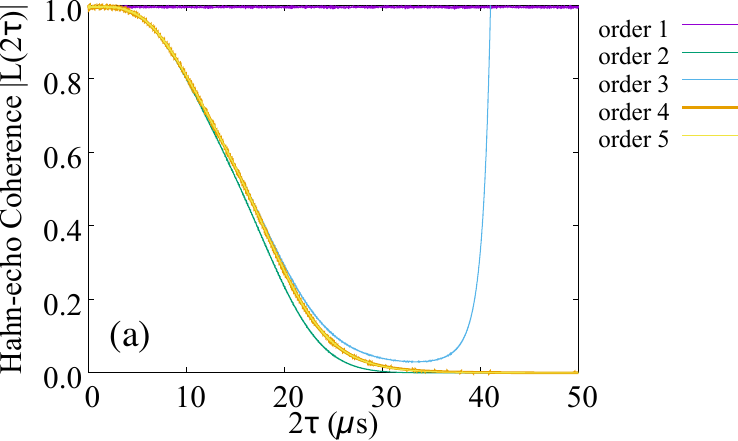}&
\includegraphics[width=0.43\textwidth]{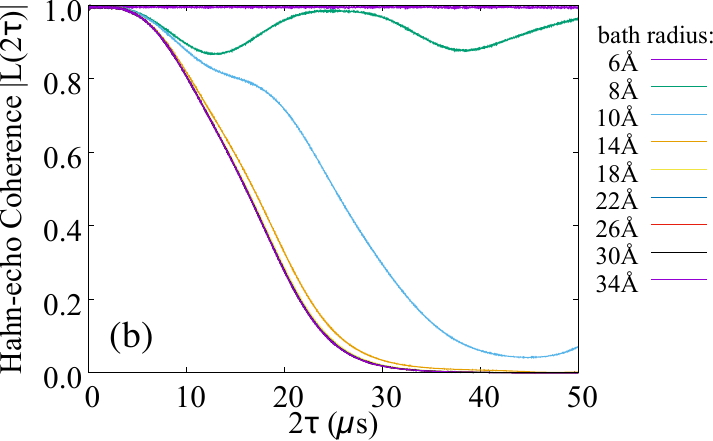}
\end{array}$

\caption{(a) Hahn-echo coherence functions for increasing CCE order. There is essentially no change going from order 4 to order 5. In this example the magnetic field is along the direction of the GNR and is $6\T$. (b) Hahn-echo coherence functions at increasing bath radius, showing convergence at around $26\ang$. The magnetic field in this example is the same as in (a).}\label{fig:convergence}
\end{figure*}

\section{Convergence tests for CCE calculations \label{ConvergenceTest}}

A convergence test of the Hahn-echo coherence function against CCE order and bath radius is shown in Fig.~\ref{fig:convergence}. This calculation is for the ground state atomic configuration as in Fig.~\ref{fig:isomeric_structures}(a) and for a $6\T$ magnetic field in the direction of the GNR. There is essentially no change in the coherence function when the CCE order is increased from 4 to 5 [Fig.~\ref{fig:convergence}(a)]. The coherence function is also convergent when the bath radius is increased to around $26\ang$ [Fig.~\ref{fig:convergence}(b)]. Tests performed with different field directions and strengths result in the same CCE order and bath radius for convergent coherence function.

\clearpage 

\begin{widetext}

\section{Coherence function of the two-spin model \label{model}}
  
The Hahn-echo coherence function of central spin for the $S=\frac{1}{2}$, $I=1$ model
of Section \ref{two-spin_model} is 

\begin{eqnarray}
L(t=2\tau)&=&\frac{1}{3\left[{A_{zx}}^{4}+2\left({A_{zz}}^{2}+4B^{2}{\gamma_{N}}^{2}\right){A_{zx}}^{2}+\left({A_{zz}}^{2}-4B^{2}{\gamma_{N}}^{2}\right)^{2}\right]^{2}} \times \nonumber\\
&&\Bigl\{ a\cos\left[\left(\frac{1}{2}\sqrt{A_{zx}^{2}+(A_{zz}+2B\gamma_{N})^{2}}\right)t\right]+b\cos\left[\left(\frac{1}{2}\sqrt{A_{zx}^{2}+(A_{zz}-2B\gamma_{N})^{2}}\right)t\right]\nonumber\\
&&+c\cos\left[\left(\sqrt{A_{zx}^{2}+(A_{zz}+2B\gamma_{N})^{2}}\right)t\right]+d\cos\left[\left(\sqrt{A_{zx}^{2}+(A_{zz}-2B\gamma_{N})^{2}}\right)t\right]\nonumber\\
&&+e\cos\left[\left(\frac{1}{2}\sqrt{A_{zx}^{2}+(A_{zz}+2B\gamma_{N})^{2}}-\sqrt{A_{zx}^{2}+(A_{zz}-2B\gamma_{N})^{2}}\right)t\right]\nonumber\\
&&+f\cos\left[\left(\sqrt{A_{zx}^{2}+(A_{zz}+2B\gamma_{N})^{2}}-\sqrt{A_{zx}^{2}+(A_{zz}-2B\gamma_{N})^{2}}\right)t\right]\nonumber\\
&&+g\cos\left[\frac{1}{2}\left(\sqrt{A_{zx}^{2}+(A_{zz}+2B\gamma_{N})^{2}}-\sqrt{A_{zx}^{2}+(A_{zz}-2B\gamma_{N})^{2}}\right)t\right]\nonumber\\
&&+h\cos\left[\left(\sqrt{A_{zx}^{2}+(A_{zz}+2B\gamma_{N})^{2}}-\frac{1}{2}\sqrt{A_{zx}^{2}+(A_{zz}-2B\gamma_{N})^{2}}\right)t\right]\nonumber\\
&&+i\cos\left[\frac{1}{2}\left(\sqrt{A_{zx}^{2}+(A_{zz}+2B\gamma_{N})^{2}}+\sqrt{A_{zx}^{2}+(A_{zz}-2B\gamma_{N})^{2}}\right)t\right]\nonumber\\
&&+j\cos\left[\left(\sqrt{A_{zx}^{2}+(A_{zz}+2B\gamma_{N})^{2}}+\frac{1}{2}\sqrt{A_{zx}^{2}+(A_{zz}-2B\gamma_{N})^{2}}\right)t\right]\nonumber\\
&&+k\cos\left[\left(\frac{1}{2}\sqrt{A_{zx}^{2}+(A_{zz}+2B\gamma_{N})^{2}}+\sqrt{A_{zx}^{2}+(A_{zz}-2B\gamma_{N})^{2}}\right)t\right]\nonumber\\
&&+l\cos\left[\left(\sqrt{A_{zx}^{2}+(A_{zz}+2B\gamma_{N})^{2}}+\sqrt{A_{zx}^{2}+(A_{zz}-2B\gamma_{N})^{2}}\right)t\right]+x\Bigr\},\label{eq:two_spin_model_L_expression}
\end{eqnarray}
where
\begin{eqnarray}
&&a=b=64A_{zx}^{2}B^{2}{\gamma_{N}}^{2}[A_{zx}^{4}+({A_{zz}}^{2}-4B^{2}{\gamma_{N}}^{2})^{2}+2A_{zx}^{2}({A_{zz}}^{2}-2B^{2}{\gamma_{N}}^{2})],\nonumber\\
\noalign{\medskip}
&&c=d=192{A_{zx}}^{4}B^{4}{\gamma_{N}}^{4},\nonumber\\
\noalign{\medskip}
&&e=h=j=k=-128{A_{zx}}^{4}B^{4}{\gamma_{N}}^{4},\nonumber\\
\noalign{\medskip}
&&f=l=32{A_{zx}}^{4}B^{4}{\gamma_{N}}^{4},\nonumber\\
\noalign{\medskip}
&&g=i=-32A_{zx}^{2}B^{2}{\gamma_{N}}^{2}(A_{zx}^{2}+{A_{zz}}^{2}-4B^{2}{\gamma_{N}}^{2})^{2},\nonumber\\
\noalign{\medskip}
&&x=3A_{zx}^{8}+3({A_{zz}}^{2}-4B^{2}{\gamma_{N}}^{2})^{4}+4A_{zx}^{6}(3{A_{zz}}^{2}-4B^{2}{\gamma_{N}}^{2}) 
 +4A_{zx}^{2}({A_{zz}}^{2}-4B^{2}{\gamma_{N}}^{2})^{2} \nonumber\\
\noalign{\smallskip}
&&\qquad\qquad \times (3{A_{zz}}^{2}-4B^{2}{\gamma_{N}}^{2})
+2A_{zx}^{4}(9{A_{zz}}^{4}-40{A_{zz}}^{2}B^{2}{\gamma_{N}}^{2}+176B^{4}{\gamma_{N}}^{4}).\label{ESEEM_coefficients}
\end{eqnarray}
The ESEEM amplitude, or modulation depth, for each frequency is just the coefficient in front of each cosine functions, and they share the same denominator ${3\left[{A_{zx}}^{4}+2({A_{zz}}^{2}+4B^{2}{\gamma_{N}}^{2}){A_{zx}}^{2}+({A_{zz}}^{2}-4B^{2}{\gamma_{N}}^{2})^{2}\right]^{2}}$.

\clearpage

\section{Product rules in the cases of magnetic fields along directions (ii) and (iii) \label{product_rules_directions_ii_and_iii}}

The product rules among different nuclear spin species and among different N nuclear spins as described in Sect.~III(B) are valid for magnetic fields in all the directions we consider in this paper. Examples of product rules for field directions (ii) and (iii) are shown in this appendix section in Figs.~\ref{fig:product_elements_y_ligand},~\ref{fig:product_elements_VO} and ~\ref{N_product_rule_direction_ii_and_iii}.

\begin{figure*}[htp]
$\begin{array}{cc}
\includegraphics[width=0.4\linewidth]{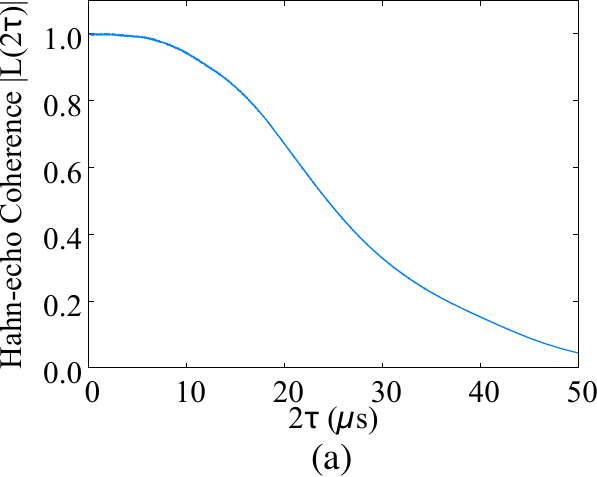}&
\includegraphics[width=0.4\linewidth]{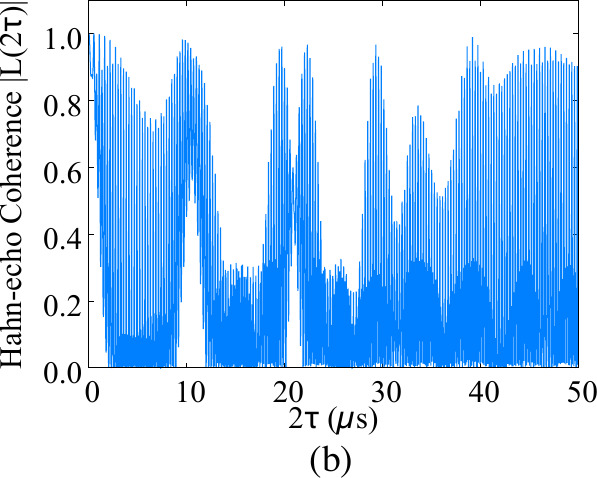}\\
\includegraphics[width=0.4\linewidth]{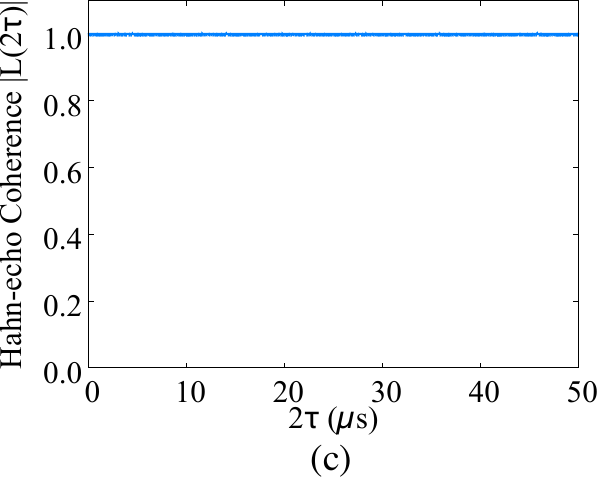}&
\includegraphics[width=0.4\linewidth]{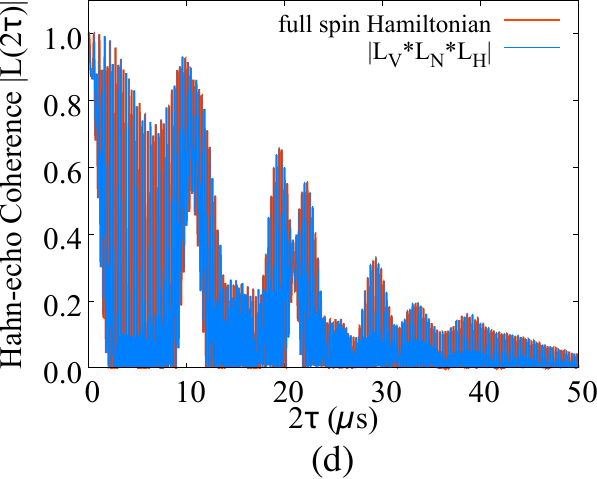}
\end{array}$
\caption{Product rule for the Hahn-echo coherence function among different elements in the VOPc@GNR system. A magnetic field of strength $1.2\T$ is applied in the direction (ii) as described in Sect. III(B). The coherence function is contributed by the central spin coupling to {(a)} H nuclei, {(b)} N nuclei, and {(c)} the V nucleus. In {(d)}, the coherence function due to all elements present in the spin Hamiltonian is seen to closely follow the product of (a),(b) and (c). }\label{fig:product_elements_y_ligand}
\end{figure*}

\begin{figure*}[htp]
$\begin{array}{cc}
\includegraphics[width=0.4\linewidth]{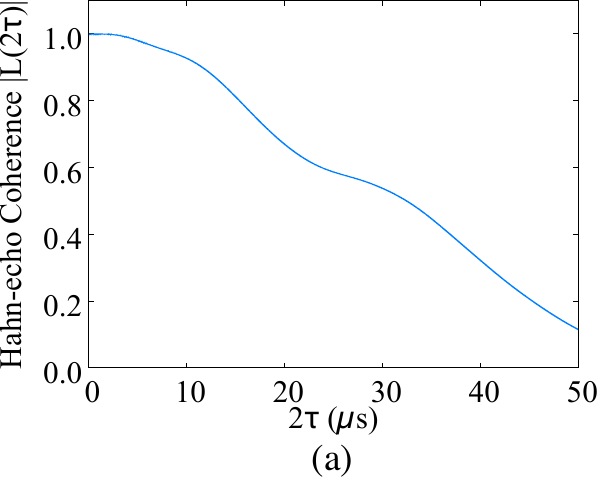}&
\includegraphics[width=0.4\linewidth]{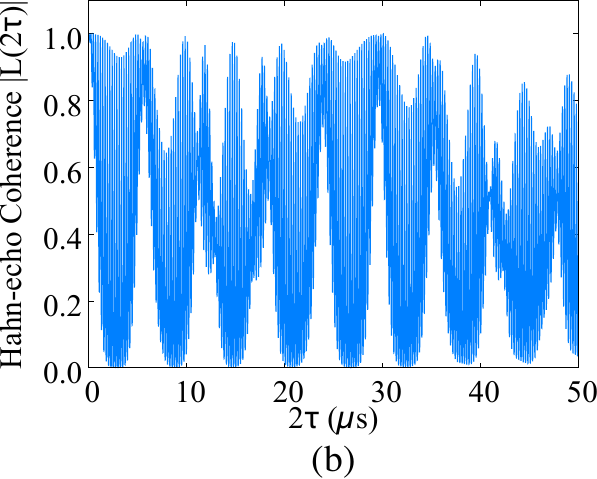}\\
\includegraphics[width=0.4\linewidth]{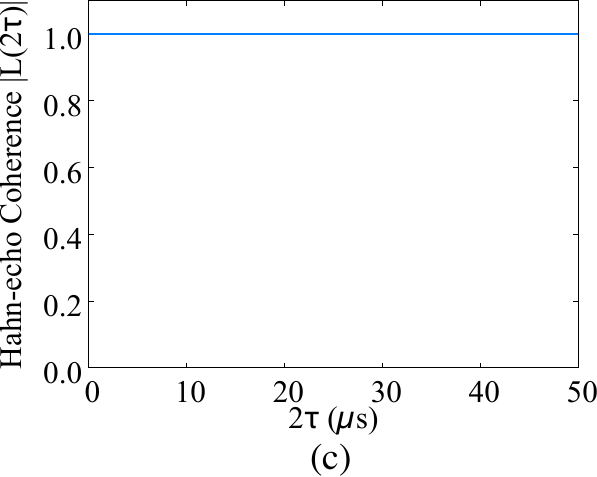}&
\includegraphics[width=0.4\linewidth]{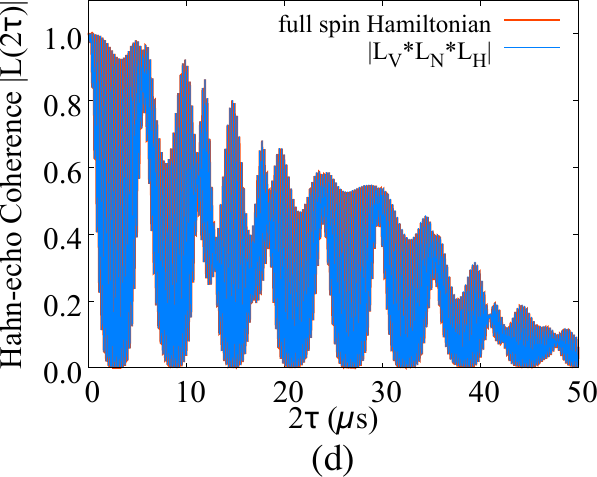}
\end{array}$
\caption{Product rule for the Hahn-echo coherence function among different elements in the VOPc@GNR system. A magnetic field of strength $1.4\T$ is applied in the direction (iii) as described in Sect. III(B). The coherence function is contributed by the central spin coupling to {(a)} H nuclei, {(b)} N nuclei, and {(c)} the V nucleus. In {(d)}, the coherence function due to all elements present in the spin Hamiltonian is seen to closely follow the product of (a),(b) and (c). }\label{fig:product_elements_VO}
\end{figure*}

\begin{figure*}[htp]
$\begin{array}{cc}
\includegraphics[width=0.4\linewidth]{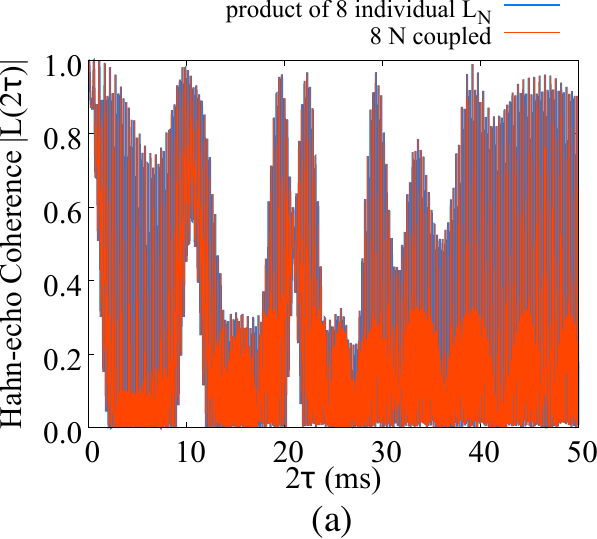}&
\includegraphics[width=0.4\linewidth]{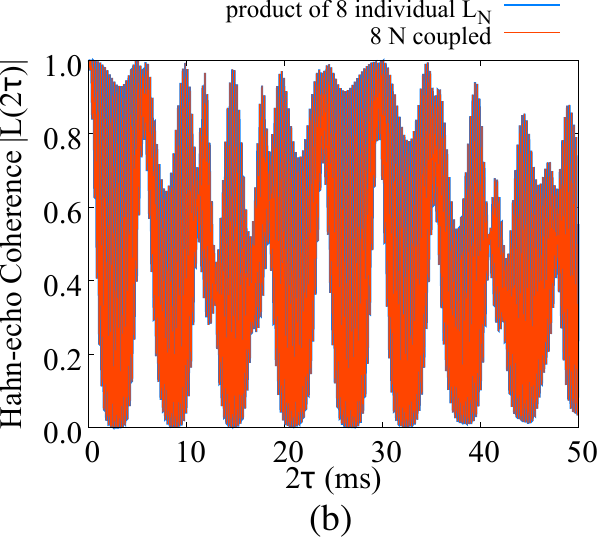}
\end{array}$
\caption{Comparison between the coherence functions due to eight N nuclear spins together and the product of coherence functions due to each individual N for {(a)} a magnetic field of strength $1.2\T$ applied in the direction (ii), {(b)} a magnetic field of strength $1.4\T$ applied in the direction (iii).}\label{N_product_rule_direction_ii_and_iii}
\end{figure*}

\end{widetext}

\end{document}